\documentclass[10pt,twocolumn,superscriptaddress,aps,prb,balancelastpage,longbibliography]{revtex4-1}

\usepackage{graphicx}
\usepackage{natbib,hyperref,ulem,soul}
\usepackage{amsmath,epsfig,color,amssymb}
\usepackage{comment}
\usepackage{enumitem}

\newcommand{\barOmega}{\ensuremath{\bar{\Omega}}}
\newcommand{\btau}{\ensuremath{\boldsymbol{\tau}}} 
 
\newcommand{\bOmega}{\ensuremath{\boldsymbol{\Omega}}} 
\newcommand{\gcrit}{\ensuremath{g_\mathrm{crit}}}

\newcommand{\tr}{\ensuremath{\mathrm{tr}}}
\newcommand{\bd}{\ensuremath{\mathbf{d}}}

\newcommand{\br}{\mathbf{r}}
\newcommand{\bA}{\mathbf{A}}
\newcommand{\bk}{\mathbf{k}}
\newcommand{\bK}{\mathbf{K}}
\newcommand{\diag}{\ensuremath{\mathrm{diag}}}

\newcommand{\hatmathO}{\hat{\mathcal{O}}}

\begin{document}

\author{Hossein Dehghani}
\affiliation{Joint Quantum Institute, College Park, 20742 MD, USA}
\affiliation{The Institute for Research in Electronics and Applied Physics, University of Maryland, College Park, 20742 MD, USA}

\author{Mohammad Hafezi}
\affiliation{Joint Quantum Institute, College Park, 20742 MD, USA}
\affiliation{The Institute for Research in Electronics and Applied Physics, University of Maryland, College Park, 20742 MD, USA}

\author{Pouyan Ghaemi}
\affiliation{Physics Department, City College of the City University of New York, New York, NY 10031, USA}
\affiliation{Physics Program, Graduate Center of the City University of New York, NY 10031, USA}

\title{Light-induced topological superconductivity via Floquet interaction engineering} 

\begin{abstract}
We propose a mechanism for light-induced unconventional superconductivity in a two-valley semiconductor with a massive Dirac type band structure. The superconducting phase results from the out-of-equilibrium excitation of carriers in the presence of Coulomb repulsion and is stabilized by coupling the driven semiconductor to a bosonic or fermionic thermal bath. We consider a circularly-polarized light pump and show that by controlling the detuning of the pump frequency relative to the band gap, different types of chiral superconductivity would be induced. The emergence of novel superconducting states, such as the chiral $p$-wave pairing, results from the Floquet engineering of the interaction. This is realized by modifying the form of the Coulomb interaction by projecting it into the states that are resonant with the pump frequency. We show that the resulting unconventional pairing in our system can host topologically protected chiral bound states. We discuss a promising experimental platform to realize our proposal and detect the signatures of the emergent superconducting state. 
\end{abstract}
\maketitle

\textit{Introduction.---} 
Possibility of generating superconductivity (SC) in periodically driven systems has been long investigated in semiconductors \cite{galitsky1970electric}, and it has been argued that under population inversion repulsive interactions can lead to a superconducting instability \cite{Kirzhnits1973Superconductivity, Galitski1974Feasibility,goldstein2015photoinduced,langenberg1986nonequilibrium}. Recent developments in non-equilibrium Floquet band engineering \cite{Oka2009Photovoltaic, Kitagawa2010,  Lindner2011Floquet, Cayssol2012, Rechtsman2013Photonic, Perez2014Floquet, HDehghani2014Dissipative, Titum2015Anomalous, Kim2020, McIver2020} has revived interest in periodically driven and light-induced interacting quantum phases of matter \cite{Hu2014Optically, Zhang2015Chiral, Mitrano2016Possible, Po2016Chiral, Sota2016Hubbard, Babadi2017Theory, Dehghani2017Dynamical,Kennes2019Light, Dehghani2020Optical}. In particular, recently such effects were studied in hexagonal semiconductors such as hexagonal Boron-Nitride or two dimensional transition metal dichalcogenides \cite{hart2019steady}. It has been proposed that light-induced non-thermal population occupation can lead to interband superconducting correlations in the presence of repulsive interactions and fermionic or bosonic baths \cite{Porta2019Feasible}. Therefore, it is intriguing to question whether more exotic form of superconductivity could be achieved in such driven systems.

\begin{figure}[ht!]
    \centering 
    \includegraphics[width=.47\textwidth]{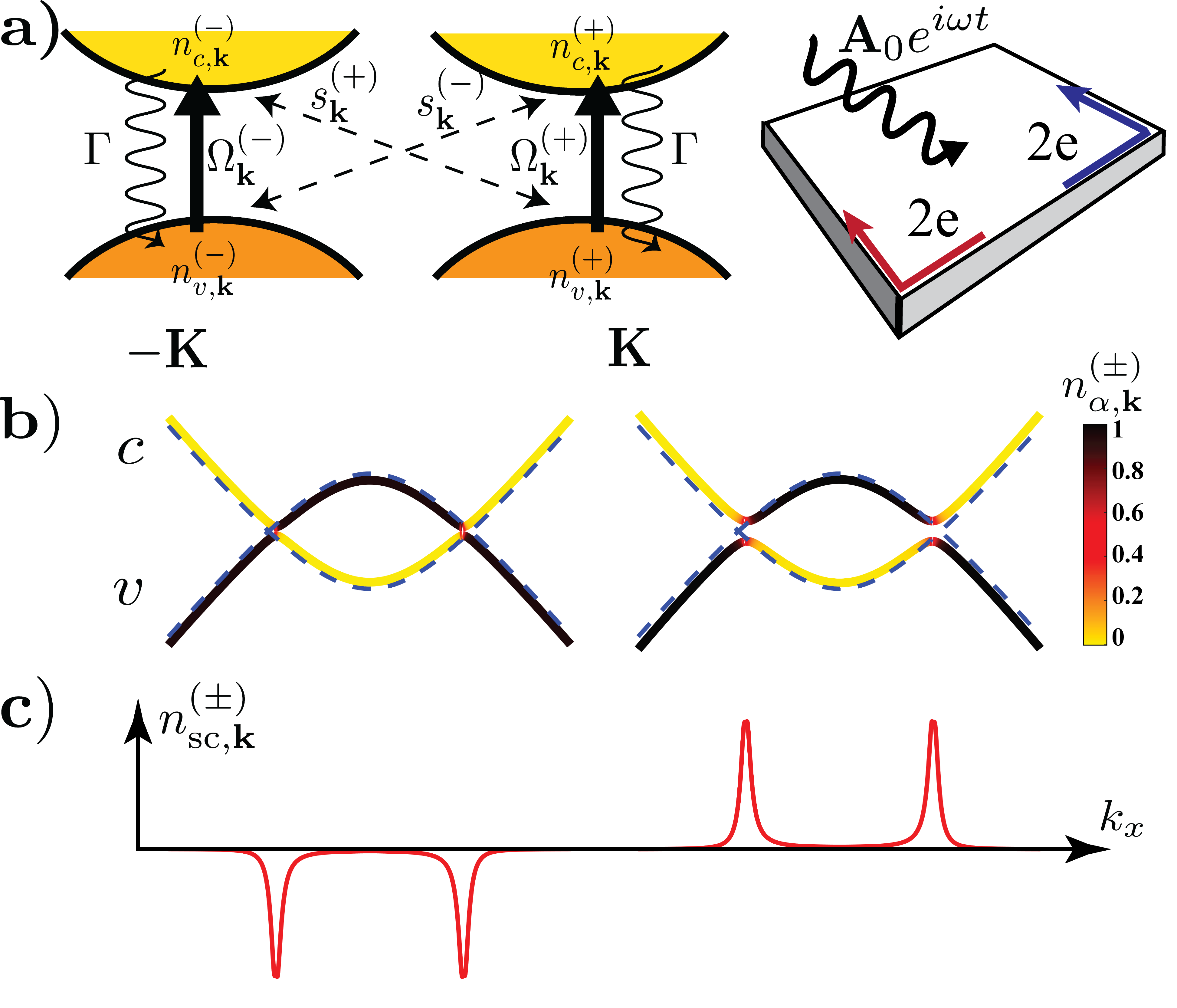}
    \caption{(a) (Left) 
    Relevant processes in our proposal in a 2D semiconductor around the two Dirac points represented by $\pm \bK$. The Rabi frequency around the $\bK$ ($-\bK$) valley, and the system-reservoir couplings are labeled by $\barOmega_{\bk}^{(+)}$ ($\barOmega^{(-)}_{\bk}$), and $\Gamma$, respectively. $n^{(\pm)}_{v,\bk}\ (n^{(\pm)}_{c,\bk})$, and $s^{(+)}_{\bk} (s^{(-)}_{\bk})$ represent the occupation probabilities of the valence (conduction) band, and the anomalous interband pairing between conduction band around valley $\bK\ (-\bK)$ and valance band around valley $-\bK\ (\bK)$, respectively. (Right) Geometry of the sample:  counter-propagating emergent superconducting edge states with energy $ \pm eA_0v/2$ depicted with blue/red colors. 
    (b) The driven (static) energy spectrum in the rotating frame plotted with a solid (dashed) line colored according to their population probabilities. (c) The interband pairing population, $n^{(\pm)}_{\mathrm{sc},\bk}=1- n^{(\pm)}_{v,\bk} - n^{(\mp)}_{c,-\bk}$, in terms of momentum.}
\label{fig:schematics}
\end{figure}

In this paper, we show that the extension of these ideas could lead to creation and manipulation of topological superconducting phases. In particular, we show that optical pumping of electrons in such two-dimensional (2D) semiconductors can generate topologically non-trivial chiral SC \cite{Read2000Paired, Qi2010Chiral} which hosts topologically protected chiral edge states in the prethermal regime of our driven system. The idea is illustrated in Fig.~\ref{fig:schematics}(a), where we apply a circularly polarized laser field, in the presence of an external bath to create the population imbalance, required for the development of a non-equilibrium superconducting phase. 

The key underlying mechanism for the development of unconventional SC in our system is the following. By varying the pump's frequency, we excite photocarriers of select momentum classes. Due to the optical valley polarization, this leads to an asymmetric occupation distribution around the resonance surfaces in the two valleys, as in Fig.~\ref{fig:schematics}(b). This non-equilibrium occupation creates an effective \textit{pairing} population inversion around one of the valleys, Fig.~\ref{fig:schematics}(c), which leads to an interband pairing of electrons for a repulsive density-density interaction, i.e., the pairing population inversion effectively changes the interaction sign. We should note that although the superconductivity due to the population inversion has not been yet observed, recent advancements in development of transient light induced quantum phases of matter, brings the realization of such forms of superconductivity much more conceivable.

To study this pairing, the bare density-density interaction should be projected into the band basis, composed of Bloch wave functions. Due to the non-zero Berry curvature of Bloch wave functions around each valley, the effective interaction has a chiral nature, and can be decomposed into different angular momentum channels where each channel has a different dependence on the momentum of electrons. Combined with the fact that the momentum distribution of the excited photocarriers are controlled by the pump's frequency, our setup allows for engineering the dominant form of electron-electron (e-e) interaction. Consequently, we find frequency regimes where a chiral p-wave pairing becomes more favorable than a $s$-wave pairing. Therefore, our results indicate that the periodic drive could be a powerful tool to not only engineer a band, but also control the form and strength of the interaction \cite{Lee2018Floquet, PhysRevLett.119.247403, cian2020engineering}.

\textit{Model.---} The system considered in this Letter consists of a 2D semi-conductor with honeycomb lattice structure, such as a single layer hexagonal Boron nitrate (h-BN) \cite{KAWAGUCHI20081171,hbn} or transition metal dichalcogenides (TMD) \cite{Wang2012,Xu2014}. The electronic band structure consists of two degenerate valleys and the broken inversion symmetry leads to a gap at two Dirac points $\textbf{K}$ and $-\bK$ at the corners of Brillouin zone (BZ) which are labeled by $\eta=(\pm)$, respectively. The semiconductor is driven by a laser beam, whose frequency is slightly larger than the semiconductor gap. The Hamiltonian describing the system is $H_s = H_K + H_{\mathrm{e-e}}$
where $H_K$ is a driven kinetic term, and $H_{\mathrm{e-e}}$ is an e-e interaction. The driven Hamiltonian for the semiconductor has the form
\begin{equation}\label{mh} 
H_K = \sum_{a, b, \eta=\pm}c_{a,\bk}^{\eta\dagger}\left[ \left(\bd^{\eta}_{\bk}+\bOmega^{\eta}(t)\right).\btau_{ab}-\mu_{\bk}\mathbf{1}_{ab}\right]c^{\eta}_{b,\bk},
\end{equation}
where $c_{a,\bk}^{\eta\dagger}$ is the electron creation operator of sublattice type $a$ in the vicinity of valley $\eta \bK$, $\tau_{i}$ with $i=\{x, y, z\}$ is the Pauli matrix acting on the sublattice space in the unit-cell, and $\mu_{\bk}$ is the chemical potential which we assume to be momentum independent. The low energy Hamiltonian of the two valleys is given by $\bd^{\eta}_{\bk} = (v k_x, \eta v k_y, m- \kappa k^2)$ where $\bk$ denotes the deviation from the $\bK$ or $-\bK$ points in the BZ, and $m$, $v$ and $\kappa$, corresponding to the band gap, Fermi velocity, and the band curvature, respectively, and $k^2 = k_x^2+k_y^2$. The optical driving of the system with a circularly polarized laser field is described by minimal coupling ($\bk \rightarrow \bk + e\bA$), where the laser field's vector potential is $\bA(t) = A_0(\cos \omega t, \sin \omega t, 0)$, with $A_0$ and $\omega$ labeling the amplitude and frequency of the pump, respectively and we set $\hbar=1$. Up to linear order in $A_0$ the associated Rabi vector of the optoical pump is,
$\bOmega^{\eta}(t)=eA_0v\notag \left( \cos \omega t, \eta \sin \omega t, -2\frac{\kappa}{v} (k_x\cos\omega t +  k_y\sin\omega t)\right)$. For simplicity, we ignore the effect of the physical spin which only affects our results when the spin-orbit coupling is comparable to the semiconductor gap \cite{goldstein2015photoinduced}. 

In the following, we denote the corresponding eigenenergies and eigenstates of the undriven Hamiltonian by $\epsilon^{\eta}_{\alpha,\bk}$ and $|u^{\eta}_{\alpha, \bk}\rangle$, where the valence and conduction bands are labeled by $\alpha = \{ v,c\}$, respectively. 

For the e-e interaction, we consider a repulsive density-density potential $U(\br-\br')$, with the corresponding Hamiltonian
\begin{align}
H_{\mathrm{e-e}} = \int d^2 \br d^2 \br' \sum_{a,b} \psi^{\dagger}_a(\br) \psi^{\dagger}_b(\br') U(\br-\br')  \psi_b(\br') \psi_a(\br), 
\label{coulombInteraction}
\end{align}
where $\psi^{\dagger}_a(\br)$ represents the electronic creation operator with the sublattice index $a$. To study the possibility of Cooper pairing between electrons, we suppose that the dominant form of the interaction is a screened Coulomb interaction \cite{Loon2018Competing}. Therefore, in passing to the momentum space, such interactions are treated as a constant coupling. Denoting the Fourier transform of Coulomb potential by $U_{\bk\bk'}$, this implies that $U_{\bk\bk'}=g/N$, where $g$ is the interaction strength and $N$ stands for the number of particles in the unit-cell. We also restrict our interactions to intra-valley scatterings such that in $U_{\bk \bk'}$, $\bk$ and $\bk'$ belong to the same valley.  

To create an effective pairing population inversion, we need a thermal bath. Our bath can have a fermionic or bosonic nature, however, here we only consider a bosonic bath composed of photons or phonons which is experimentally more feasible and leave the study of the fermionic bath to Appendix 2B \cite{App}. Specifically, here we assume that our bath 
can induce relaxation processes between the valence and conduction bands via absorption/emission of photons. 

\textit{Master equation.---} To examine the out-of-equilibrium nature of SC in the presence of a thermal reservoir at temperature $T$, we use the master equation approach.
Assuming that the system-bath coupling is sufficiently weak and the bath has a short auto-correlation time, we employ the Born-Markov approximation to trace out photons from the equations of motion (EOM). We also consider large pump frequencies compared to Rabi frequencies, which allows us to use the rotating wave approximation (RWA). As a result, we obtain an effective static master equation for the density matrix of the system
$\rho_s$ \cite{Breuer2002Theory}, 
\begin{align}
    &\partial_t\rho_{s}(t) = -i [H_{s}, \rho_{s} ] \notag \\& + \sum_{\alpha={v,c}} \Gamma_{\alpha}\Big( L_{\alpha, \bk} \rho_s L_{\alpha, \bk}^{\dagger} - \frac{1}{2}\{L_{\alpha, \bk}^{\dagger}L_{\alpha, \bk},\rho_s\} \Big)\label{rhoEOM}.
\end{align}
where the dissipator operators are $L_{v, \bk} =L^{\dagger}_{c, \bk} = c^{\dagger}_{c, \bk}c_{v, \bk}$. Associated with these dissipators we suppose momentum independent decay rates $\Gamma_{\alpha}$ corresponding to an effective population $n_B$ such that $\Gamma_v =\Gamma n_B$, and $\Gamma_c = \Gamma (1+n_B)$
\cite{supp}.

\textit{Rotating Frame Transformation.---} 
By applying the RWA to the time dependent term $\bOmega(t)$ in Eq.~\eqref{mh}, this term in the rotating frame is replaced by the static vectors $\bar{\bOmega}_{\bk}^{(+)}$, and $\bar{\bOmega}_{\bk}^{(-)}$ around the $\bK$ and $-\bK$ Dirac points, whose magnitudes are respectively given by $|\bar{\bOmega}_{\bk}^{(+)}|\simeq eA_0 v$ and $|\bar{\bOmega}_{\bk}^{(-)}|\simeq eA_0 v^3k^2/(4m^2)$ \cite{supp}.  
 
The modification of the e-e interactions becomes transparent through a mode decomposition of the field operator $\psi_a(\br) = \sum_{\alpha= \{v,c \};\eta, \bk} \frac{1}{\sqrt{S}} u^{\eta, a}_{\alpha,\bk} e^{i (\eta\bK+\bk).\br} c^{\eta}_{\alpha, \bk}$ in Eq.~\eqref{coulombInteraction}, where $S$ is the quantization area. The resulting projected Hamiltonian has a contribution in the Cooper channel for the interband pairing as, \footnote{There are additional contributions in the Cooper channel which are proportional to the overlap of the valence and conduction Bloch wave functions at close momenta. Due to the orthogonality of the valence and conduction band eigenstates these contributions are negligible.} 
\begin{align}
H_{\mathrm{e-e}} = \sum_{\bk, \bk', \eta=\pm}  \bar{U}_{\bk\bk'} c^{\eta\dagger}_{v, \bk} c^{-\eta\dagger}_{c, -\bk} c^{\eta}_{c,\bk'} c^{-\eta}_{v,-\bk'},    
\end{align}
where the projected density-density interaction is
$\bar{U}_{\bk\bk'}=(g/N) \langle u^{\eta}_{v,\bk} | u^{\eta}_{v,\bk'} \rangle \langle u^{-\eta}_{c,-\bk} | u^{-\eta}_{c,-\bk'} \rangle.$
We see that the Bloch wave functions which encompass the topological characteristics of the system, control the form of e-e interactions. 
The crucial effect of the Berry curvature of the band structure on the e-e interactions is embedded in the Bloch wave function overlaps in $\bar{U}_{\bk\bk'}$ 
which after inserting for the eigenstates can be decoupled in three channels according to their angular momenta
$\langle u^{\eta}_{v,\bk} | u^{\eta}_{v,\bk'} \rangle \langle u^{-\eta}_{c,-\bk} | u^{-\eta}_{c,-\bk'} \rangle = \sum_{l=0,1,2} f^{(l)}_{\bk}f^{(l)}_{\bk'} e^{-il(\phi_{\bk} - \phi_{\bk'})}$
where $f^{(0)}_{\bk} = \left(1+ d_{z,\bk}/d_{\bk} \right)/2$,  $f^{(1)}_{\bk} = vk/d_{\bk}$, $f^{(2)}_{\bk} = v^2 k^2/\big(2d_{\bk}(d_{\bk}+d_{z,\bk})\big)$ and $d_{\bk} = |\textbf{d}_{\bk}|$ \cite{supp}. Correspondingly, for momenta close to the corners of BZ, 
for small $\bk$ and $\bk'$ this gives, 
$\bar{U}_{\bk\bk'} \simeq g \left(1 + \frac{F}{4} (2  \bk.\bk' -2\hat{z}.\bk\times
   \bk' -  k^2 -  k^{'2} ) \right)/N,$
where $F = v^2/m^2$ is the Berry curvature at the Dirac point. 
Recently, such topologically-induced 
modifications of the e-e interaction has been associated with modification of the excitons' spectrum ~\cite{PhysRevLett.115.166802, PhysRevLett.115.166803}.

\textit{Mean-Field analysis.---} To study the possibility of the Cooper pair condensation, we use a mean-field (MF) approximation for the e-e interaction and express it as, 
\begin{align}
&H_{\mathrm{e-e}}= -\sum_{\eta=\pm;\bk, \bk'} \Delta_{\bk}^{\eta *} \bar{U}_{\bk \bk'}^{-1}\Delta_{\bk'}^{\eta} \notag \\ &+ \sum_{\eta=\pm;\bk} \left( \Delta_{\bk}^{\eta *} c^{-\eta }_{c,-\bk} c^{\eta }_{v,\bk} + h.c. \right).\label{MeanField}
\end{align}
Notice that we introduce two pairing order parameters $\Delta_{\bk}^{(\pm)}$, depending on whether the valence electrons around the $\bK$ valley and conduction electrons around $-\bK$ valley are bound to each other or vice versa. 
Employing the MF expression above, we can use Eq.~\eqref{rhoEOM} to write a closed set of equations for the occupation numbers $n^{\eta}_{\alpha, \textbf{k}} = \tr(\rho_s c^{\eta\dagger}_{\alpha , \bk} c^{\eta}_{\alpha,\bk} )$, the polarization $p^{\eta}_{\bk}=\tr(\rho_s c^{\eta \dagger}_{c, \bk} c^{\eta}_{v,\bk})$, and the anomalous pairing $s^{\eta}_{\bk}= \tr(\rho_s c^{-\eta}_{c,-\textbf{k}}c^{\eta}_{v, \textbf{k}})$. This approach leads to legitimate results at the onset of the SC phase transition, where the distinction between the Bogoliubov quasi-particles and electrons is negligible. The EOM for $n_{\alpha, \bk}^{\eta}$ and $p_{\bk}^{\eta}$ in the absence of pairing are familiar and known as the optical Bloch equations in the literature \cite{scully_zubairy_1997, yamamoto1999mesoscopic} and we leave their derivation to the Appendix 2. Here, we only present the EOM of the anomalous pairing which is less familiar, 
\begin{align} 
\partial_t s_{\bk}^{\eta} = -i \epsilon_{t,\bk}s_{\bk}^{\eta} - i \Delta^{\eta}_{\bk}(1- n_{v,\bk}^{\eta} - n_{c,-\bk}^{-\eta}) - \frac{1}{2}\Gamma_{t} s_{\bk}^{\eta}, \label{rotatedEOMs}
\end{align}
where we define the total decay rate as $\Gamma_{t} = \Gamma_{v} + \Gamma_{c}$, and the total kinetic energy as $\epsilon_{t,\bk} = \epsilon_{v,\bk} +\epsilon_{c,\bk}$. We note that on the right-side of this equation, the two occupation probabilities in the parenthesis belong to two different valleys which is a manifestation of the Cooper pairing.
In the steady state where the right-hand side of Eq.~\eqref{rotatedEOMs} vanishes, $s_{\bk}^{\eta}$ satisfies $s_{\bk}^{\eta} = -i \Delta^{\eta}_{\bk}n^{\eta}_{\mathrm{sc},\bk}/(i\epsilon_{t,\bk}+\Gamma_t/2)$ where we define the \textit{interband} pairing population as $n^{\eta}_{\mathrm{sc},\bk}= 1- n^{\eta}_{v,\bk} - n^{-\eta}_{c,-\bk}$. Since $n^{\eta}_{v,\bk}$ and $n^{\eta}_{c,-\bk}$ can be independently populated due to the optical valley selection rules, under non-equilibrium conditions the pairing population can acquire a finite value. 

Using the MF definition of the anomalous pairing $s_{\bk}$, the self-consistency equation at the onset of the phase transition becomes,
$\Delta_{\bk}^{\eta} \simeq - \sum_{\bk'}\bar{U}_{\bk \bk'} s_{\bk'}^{\eta}$. However, since in dissipative systems, $s_{\bk'}$  may have a finite imaginary part, this equation should be modified by a more accurate Keldysh approach \cite{goldstein2015photoinduced, hart2019steady}. For  weak decay rates, we get, $ \Delta_{\bk}^{\eta} \simeq - \sum_{\bk'}\bar{U}_{\bk \bk'}\mathrm{Re}\left[s_{\bk'}^{\eta}/\Delta_{\bk'}^{\eta} \right]$ where the real part bracket indicates the dissipative suppression of the pairing and ensures that $\Delta_{\bk}^{\eta}$ remains real. Correspondingly, in the steady-state this equation gives, 
\begin{align}
\Delta_{\bk}^{\eta} = -\sum_{\bk'} \bar{U}_{\bk \bk'}\frac{\epsilon_{t,\bk}}{\epsilon_{t,\bk}^{2} + \Gamma_t^{2}} n_{\mathrm{sc}, \bk'}^{\eta}\Delta_{\bk'}^{\eta},\label{gapEqnK}
\end{align}
where we remark that $n_{\mathrm{sc}, \bk'}^{\eta}$ effectively determines the interaction sign \cite{goldstein2015photoinduced,hart2019steady}. 
The asymmetry of the Rabi frequencies at the two valleys results in the corresponding steady state occupation probabilities to differ significantly as depicted in Fig.~\ref{fig:schematics}(b). 
For the polarization we have chosen, this leads to a positive (negative) value for $n_{\mathrm{sc}, \bk}^{(+)}$ ($n_{\mathrm{sc}, \bk}^{(-)}$) as it is displayed in Fig.~\ref{fig:schematics}(c). Thus, with a repulsive interaction, we can have a SC instability by developing a non-vanishing value for the order parameter $\Delta^{(-)}_{\bk}$, while $\Delta^{(+)}_{\bk}$ remains vanishing. Thus, in what follows we drop the valley superscript in $\Delta_{\bk}$ for the non-vanishing order parameter. 

We also highlight that the form of $\bar{U}_{\bk \bk'}$ crucially determines the form of the resulting gap. In fact, the self-consistency equation above can be solved using a simple ansatz for the gap-function of the form,
\begin{align}
\Delta^{(l)}_{\bk} = e^{-i l \phi_{\bk}} f_{\bk}^{(l)}\Delta^{(l)}.
\end{align}
where $l=\{0,1,2\}$ should be ascribed to the angular momentum of the $s, p$ and $d$-wave pairing modes and $f^{(l)}$'s play the role of the SC form factors. 
Using this ansatz and inserting $\epsilon_{t,\bk}=2\mu$, the linearized gap equations for the three different types of pairing decouple and the critical coupling strength for each channel reads,
\begin{align}
    &\frac{1}{g^{(l)}_{\mathrm{crit}}} = \frac{1}{N}\sum_{\bk'}  f^{(l)2}_{\bk'} \frac{2 \mu}{4\mu^{2} + \Gamma_t^{2}  } \Big(\frac{1}{2\zeta^{(-)}_{-\bk'}+1} - \frac{1}{2\zeta^{(+)}_{\bk'}+1} \Big), \label{gcrit}
\end{align}
where $\zeta^{\eta}_{\bk} = (\barOmega^{\eta 2}_{\bk, x} + \barOmega^{\eta 2}_{\bk, y})/(\epsilon^{2}_{d,\bk} + \frac{1}{4}\Gamma_t^{2} )$, and the detuning frequency is $\epsilon_{d,\bk} =\epsilon_{c,\bk} - \epsilon_{v,\bk}  - \omega$. Since we only consider direct optical transitions, $\zeta^{\eta}_{\bk}$ is essentially the pairing population in each momentum class, in the weak-drive limit.

\begin{figure} 
\includegraphics[width=.48\textwidth]{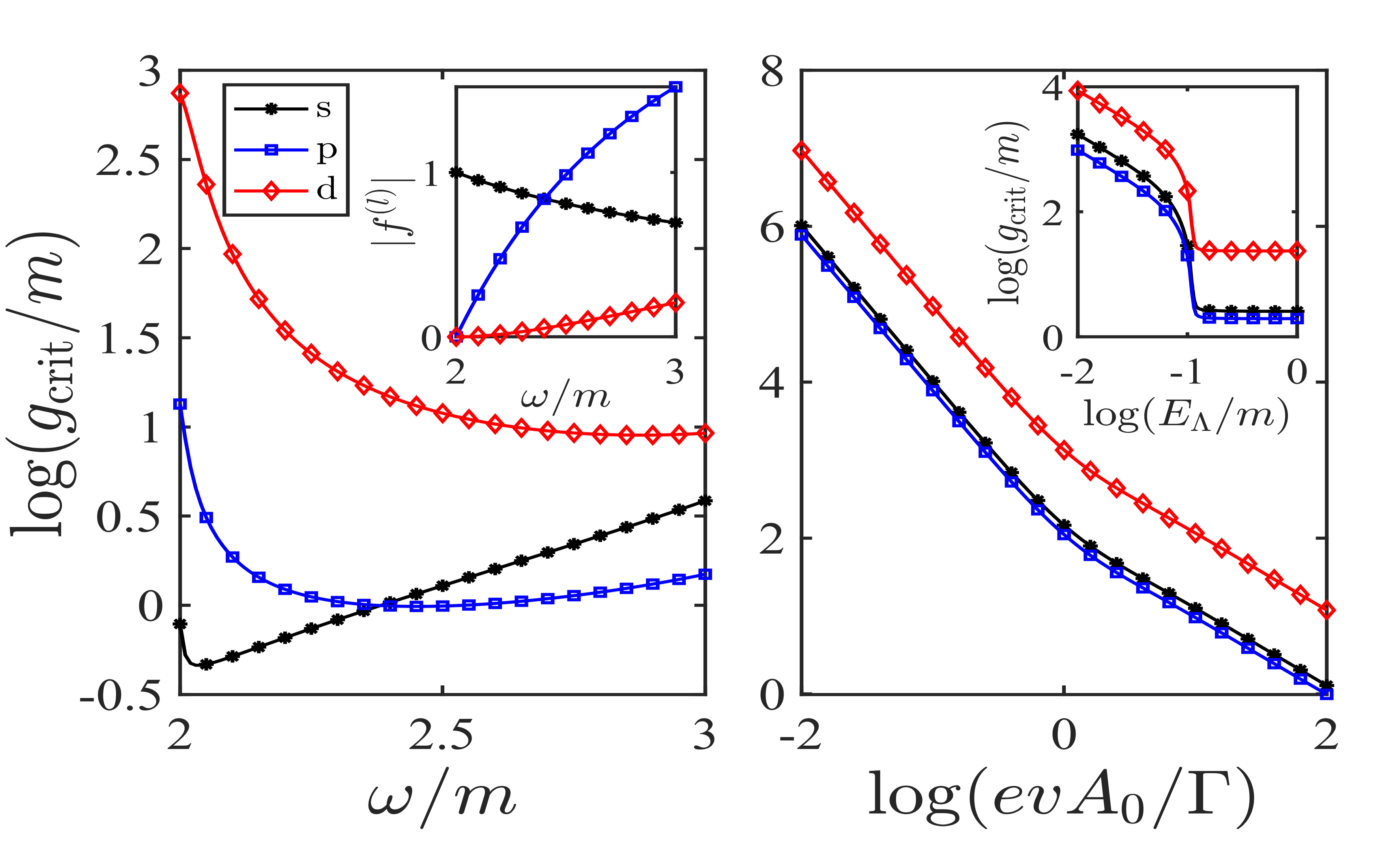}
\caption{Critical coupling, $g_{\mathrm{crit}}$, for development of superconductivity in three $s, p$ and $d$ wave channels. We choose $\Gamma/m=0.0002$, $v/(ma)=0.4$, $\kappa/(ma^2) = 0.04$, $\mu/m=0.01$, $n_B=0.001$ where $a$ is the lattice constant (a) $g_{\mathrm{crit}}$ vs drive's frequency with $e v A_0/m = 0.02$ and $E_{\Lambda}/m=0.5$. Inset: $f^{(l)}$ form factors as a function of the resonant frequency. (b) $g_{\mathrm{crit}}$ vs drive's amplitude: we choose $\omega/m = 2.5$. Inset: $g_{\mathrm{crit}}$ as a function of the UV energy cutoff with the drive amplitude $e v A_0/m = 0.2$.}
\label{fig:gcrit}
\end{figure}
As illustrated in Fig.~\ref{fig:schematics}(c), the main contribution to $n_{\mathrm{sc},\bk}$ comes from around the resonance energy ring, denoted by $\bk_{r}$, where the detuning frequency vanishes $\epsilon_{d,\bk_r}= 0$. 
Hence, we consider a given UV cutoff $E_{\Lambda}$ around this surface and show that the resulting phenomena are independent of the exact value of this parameter. 

The frequency of the pump determines which momentum classes are resonantly excited. Eq.~\eqref{gapEqnK} indicates that only the states with negative $n_{\mathrm{sc}, \bk}$ can form pairing. Moreover, since the projected Coulomb interaction has momentum dependence, the critical value of the coupling strength depends on the pump frequency. This behavior is depicted in Fig.~\ref{fig:gcrit}(a) where we notice that the preferred form of pairing transforms from $s$-wave to $p$-wave as the drive frequency increases. This transition is associated with the momentum dependence of the SC form factors. Since, $n_{\mathrm{sc},\bk}$ is peaked around the resonant surface, we only need to consider the momentum-dependence of the form factors around this surface. Consequently, for a given frequency the radius of the resonance ring, $k_r(\omega)$, is obtained which can be inserted to determine the form factors, $f^{(l)}(\bk_r(\omega))$. These functions are displayed in the inset of Fig.~\ref{fig:gcrit}(a). Here, we observe a similar behavior as in the main plot of Fig.~\ref{fig:gcrit}(a), where by increasing the frequency the initially dominant $s$-wave form factor becomes subdominant with respect to the $p$-wave form factor. 

Other than the pump's frequency, critical coupling depends on the amplitude of the pump, too. This is displayed in Fig.~\ref{fig:gcrit}(b), where the horizontal axis has been chosen to be the dimensionless parameter $evA_0/\Gamma$ which appears in the gap equation through $\zeta_{\bk}$. We notice that 
the critical coupling strength of all the three SC modes always decreases as the pump amplitude increases. Specifically, in the low-intensity limit $(evA_0 \lesssim \Gamma$), the critical coupling is inversely proportional to the intensity $\propto(\frac{\Gamma}{evA_0})^2$ which could be associated with the fact that the peak value of excited population ($|n^{(\pm)}_{\mathrm{sc},\bk}|\simeq \zeta^{(+)}_{\bk}$) increases linearly with the intensity.   At higher intensities, where $evA_0 \gtrsim \Gamma$, this behavior changes, since the width of $n^{(\pm)}_{\mathrm{sc}, \bk}$, i.e., the number of momentum classes participating in pairing, keeps increasing and therefore we do not observe a saturating behavior. 
Furthermore, in the inset of Fig.~\ref{fig:gcrit}(b),  we depict $\gcrit$ as a function of the energy cut-off which shows that once $E_{\Lambda}$ becomes comparable to the band gap, the cutoff dependence of $\gcrit^{-1}$ becomes insignificant. Finally, note that $g_{\mathrm{crit}}$ can be controlled directly by the chemical potential and its minimum value is reached when $\mu = \Gamma_t/2$. This feature provides a high tunability in choosing the other parameters of our system 
 \cite{supp}.

\textit{Signatures of topological SC.}
Given the diversity of proposals to realize topological phases in driven-dissipative systems, it is crucial to present the experimental signature of the light-induced topological phase in our proposal. Here, we show that the \textit{interband} p-wave pairing hosts edge supercurrents which are experimentally detectable.
To analyze the edge modes, the SC behavior of the system can be described by an effective MF Hamiltonian which includes a kinetic energy and a SC pairing between the valence(conduction) electrons at $\bK$ ($-\bK$). Due to the non-equilibrium nature of superconductivity in our system, only electrons around the resonance ring participate in SC. Hence, the steady state Hamiltonian corresponds to electrons with momentum close to the resonance surface with momenta $\bk \approx \bk_r + \delta \bk$ where $\delta \bk \ll \bk_r$ denotes the momentum deviation from the resonance surface. Thus, the effective MF Hamiltonian should be projected into these states. This structure is comparable to conventional superconductors where electrons at the Fermi energy control the properties of the superconductor and edge modes \cite{PhysRevLett.72.1526,Ghaemi2011}. To examine the existence of localized states at a hard boundary parallel to $y$ direction, we set $k_y=0$ and replace the momentum deviation from the resonant point, $\delta k_x$, with $-i \partial_x$. The resulting Hamiltonian becomes, 
\begin{equation}\label{eff}\begin{split}
    \mathcal{H}_{\mathrm{eff}} \approx & \frac{evA_0}{2}\sigma_z\tau_z  + \frac{\Delta_{\bk}}{2} \sigma_x \tau_x \\ &+\mu \sigma_0 \tau_z + \left(1 - \frac{v^2 k_r^2}{2m^2}  -  \frac{v^2 k_{r,x}}{m^2} i\partial_x \right)\sigma_z\tau_0.
\end{split}\end{equation}
where the SC order has a p-wave structure ($\Delta_{k_r}=\Delta_0 k_r$). The effective Hamiltonian commutes with $\sigma_z\tau_z$ and consequently, the eigenstates are in two independent sectors corresponding to $\sigma_z\tau_z=\pm 1$. We put the boundary at $x=0$ and the sample is in $x<0$ region. The two chiral states resulting from momenta close to each resonant momentum have energies $E_{\pm}=\pm \frac{eA_0 v}{2}$ and are separated with a gap of roughly $\Delta_0$ from the continuum states close to the resonant momenta. Their corresponding wave-functions are:
\begin{equation}
    |\Psi_{\pm}\rangle=e^{\lambda x} \frac{\sin(q_{\pm} x)}{\sqrt{2}}\left(|1\rangle_{\sigma_z}|\pm 1\rangle_{\tau_z}-i|-1\rangle_{\sigma_z}|\mp 1\rangle_{\tau_z}\right)
\end{equation}
where $\lambda=\frac{m^2}{ev^3 A_0} \frac{\Delta_{k_r}}{k_r}$, $q_\pm=\frac{\pm\mu + (1-v^2/2m^2)evA_0}{|k_r|evA_0 v^2/m^2}$. Spinors $|s\rangle_{\sigma_z(\tau_z)}$ correspond to eignspinors of $\sigma_z(\tau_z)$ Pauli matrices with eignvalue $s$. The sign $\pm$ correspond to the two sectors with $\sigma_z\tau_z=\pm 1$. It is noteworthy that these edge state have opposite chirality in the two sectors as shown in Fig.\ref{fig:schematics}(a), and can solely emerge when the pairing has a p-wave structure. Furthermore, these states carry supercurrents with no potential drop which distinguishes them from normal edge currents which was experimentally detected before in graphene \cite{McIver2020}. To verify the development of the edge states outlined about, we can use the method used before to distinguish the edge and bulk super-currents in steady-state superconducting phase \cite{Wang534}. To this end, we consider a TMD sample where part of the sample (including part of the edge) is illuminated by an appropriate laser beam. We then measure the super-current carried through the edge which is illuminated by the laser field. It is also shown that in a magnetic field, fluxoid quantization generates a periodic modulation of the edge condensate which is observable as a “fast-mode” oscillation of the critical current versus the magnetic field. It should be noted that the fast-mode frequency is distinct from the conventional Fraunhofer oscillation displayed by the bulk supercurrent and the frequency of such oscillations should increase with the superconducting area  \cite{Wang534}. Interestingly, in our setup the superconducting area can be readily increased by the increase of the laser beam width which makes such measurements convenient for our setup.

The edge states' supercurrent is contrasted with the bulk states' current through its dependence on the size of superconducting region and an external magnetic field. 
Such measurements have been realized through experiments on samples with different sizes \cite{Wang534} which in our setup is simply controlled by the width of the pump beam.

\textit{Experimental feasibility.}
Finally, we provide an estimate for the pump's amplitude based on typical energy scales in 2D two-valley semiconductors. Specifically, to verify the feasibility of the realization $p$-wave SC in our model, we need to estimate the required critical coupling constant. The promising 2D semiconductor to realize the phenomena outlined here are h-BN or TMDs with the band gap of the order of $5$ eV \cite{Elias2019, PhysRevLett.105.136805}. In these materials the screened Coulomb interaction $g$ typically is comparable to the band gap of these materials \cite{Loon2018Competing},  or can be enhanced via Coulomb engineering \cite{Raja2017Coulomb}. From Fig.~\ref{fig:gcrit}(b), we deduce that to obtain $log(g_{\mathrm{crit}}/m)\simeq 0$, one requires an electric field such that the ratio $evA_0/\Gamma$ would be of the order $10^2$. Since, typically the inverse decay rate is of the order of picoseconds \cite{Kozawa2014}, this implies that the required Rabi frequency for our proposal should be $10$ THz. For a typical semiconductor, this Rabi frequnecy corresponds to an electric field of $5\times 10^6$ V/m. Recently, strong fields with an electric field of $4\times 10^{7}$ V/m have been used in generating light-induced Hall effect in graphene \cite{McIver2020}, therefore, we believe our proposal is within the experimental reach. 

\textit{Outlook.} The Floquet engineering of the interaction described here is a versatile effect which can be generalized to other lattice symmetries where the band structure has a non-trivial topological structure.
Correspondingly, interesting directions to explore in the future are the study of similar effects in multi-layer twisted semiconductors, and the generation of other interacting topological phases. While here we suffice to studying the prethermal regime of the system which at high frequencies considered in our proposal can last for many periods of the drive \cite{Abanin2015Exponentially, Mori2016Rigorous}, for longer time scales scattering processes with acoustic phonons which mediate momentum and tend to thermalize the system, 
become important. It is fascinating to explore under what circumstances the superconducting state proposed here survives in the presence of such processes \cite{Seetharam2015Controlled, Seetharam2019Steady}. Finally, for large decay rates a mean-field approach is not satisfactory and the effects of the fluctuations should be investigated \cite{Mitra2008Dissipative, Szymanska2006Nonequilibrium,yang2020intrinsic}.

\begin{acknowledgments}
\textit{Acknowledgements.---} We would like to thank Areg Ghazaryan, Hwanmun Kim, Victor Galtiski, Vinod Menon, Sunil Mittal and Mark Rudner for illuminating discussions. HD, and MH were supported by AFOSR FA9550-19-1-0399, ARO W911NF2010232, and Simons Foundation. HD, MH thanks the hospitality of the Kavli Institute for Theoretical Physics, supported by NSF PHY-1748958. PG acknowledges support from National Science Foundation Award DMR-1824265. 
\end{acknowledgments}

\begin{widetext}

\section*{Appendix}

\setcounter{equation}{0}
\renewcommand{\theequation}{A.\arabic{equation}}

\setcounter{figure}{0}

\subsection{Rotating Wave Transformation}  
\setcounter{equation}{0}

The equations of motion (EOM) in our derivation are solved in the rotating frame. Here, we mention how we derive the required rotation. We first consider a generic traceless two-by-two Hamiltonian  $h_{\bk} = \bd^{\eta}_{\bk}.\btau$ where $\tau_i$ with $i=\{x, y, z\}$ are the Pauli matrices. The eigenstates of this Hamiltonian up to a some phase factors are given by,  
\begin{align}
    |u^{\eta}_{c,\bk}\rangle = \frac{1}{\left(2d_{\bk}(d_{\bk}+d_{z,\bk})\right)^{1/2}}\begin{pmatrix}
    d_{\bk}+d_{z,\bk}\\ d^{\eta}_{+,\bk}
    \end{pmatrix}, \quad 
    |u^{\eta}_{v,\bk}\rangle= \frac{1}{\left(2d_{\bk}(d_{\bk}+d_{z,\bk})\right)^{1/2}}\begin{pmatrix}
    d^{\eta}_{-,\bk} \\ -(d_{\bk}+d_{z,\bk}) 
    \end{pmatrix},
\end{align} 
where we have dropped the valley index in $d_{\bk}$ and $d_{z,\bk}$ which have the same form in the two valleys and defined $d^{\eta}_{\pm, \bk} = d^{\eta}_{\bk, x}\pm i d^{\eta}_{\bk, y}$. 

To apply the rotating wave approximation to a non-diagonal Hamiltonian, we need to first transform the Hamiltonian into the energy basis where it is diagonal and then apply a time-dependent rotation to the two energy levels so that the time dependence of the two transformed eigenstates becomes approximately the same. The first transformation is done through a similarity transformation by the unitary matrix $U^{\eta}_{\bk} = \begin{pmatrix} |u^{\eta}_{v,\bk}\rangle &|u^{\eta}_{c,\bk}\rangle  \end{pmatrix}$, where we have used the eigenstates of the undriven Hamiltonian, and the second transformation is realized by the diagonal time-dependent transformation $\diag(e^{i \omega t/2}, e^{-i \omega t/2})$. The combination of these two transformation is $R^{\eta}_{\bk}(t) = \begin{pmatrix} |u^{\eta}_{v,\bk}\rangle e^{i\omega t/2} &|u^{\eta}_{c,\bk}\rangle e^{-i\omega t/2} \end{pmatrix}$. Therefore, denoting the electronic spinors in the lab frame and the rotating frame via $\Psi_{\bk}^T \equiv \left(c^{\eta}_{a,\bk}, c^{\eta}_{b,\bk}\right)$, and $\tilde{\Psi}_{\bk}^T \equiv \left(\tilde{c}^{\eta}_{v,\bk},\tilde{c}^{\eta}_{c,\bk}\right)$, respectively, we have $\Psi^{\eta}_{\bk}=R^{\eta}_{\bk}(t)\tilde{\Psi}^{\eta}_{\bk}$. We apply this transformation to all of the terms in the Hamiltonian system. While the undriven Hamiltonian is trivially diagonalized, the pump Hamiltonian should be obtained after averaging over time. To evaluate the temporal average of the drive term it would be convenient to decompose the time dependent terms as,
\begin{align}
\bOmega(t)= \bOmega_c \cos(\omega t)+\bOmega^{\eta}_s \sin(\omega t).
\end{align}
where $\bOmega_c = eA_0v(1, 0, -2\frac{\kappa}{v}k_x)$ and $ \bOmega^{\eta}_s = eA_0v(0, \eta, -2\frac{\kappa}{v}k_y).$ Correspondingly, the expression that must be averaged over time is $R^{\eta\dagger}_{\bk}(t)\bOmega^{\eta}(t)R^{\eta}_{\bk}(t)$. The final result of this calculation becomes, 
\begin{align}\label{rwOmega}
    & \bar{\bOmega}_{\bk}^{(+)} = e A_0v  \left(1 + \frac{\kappa k^2}{m} - \frac{v^2 k^2}{4m^2}, 0, 0  \right), 
\end{align}
and 
\begin{align}\label{rwOmegaTilde}
& \bar{\bOmega}_{\bk}^{(-)} =   e A_0 v\left( \frac{v^2(k_y^2 - k_x^2)}{4m^2}, -\frac{v^2 k_x k_y}{2m^2}, 0\right). 
\end{align}

Similarly, we need to transform electron-electron interaction term by rotating the electronic creation and annihilation operators and averaging over time. The final result of this calculation, in addition to the right-hand side of Eq. (8) in the main text, has other contributions which include the overlap of the valence and conduction wave functions at close momenta which makes such terms negligible. 

Here, we mention that in order to integrate the gap equation, we consider an energy cutoff with respect to the resonance surface. The resonance ring in the BZ is defined by $\omega = 2 d_{\bk=\bk_{r}}$ demanding that, 
\begin{align}
    k_{r} = \frac{1}{2}\sqrt{\frac{\omega^2 -4 m^2}{v^2 - 2m\kappa}}, \label{kres}
\end{align}
where $k_{r}$ is the radius of the resonance ring. We can also use the above equation to define the integral bounds of the radial momentum through the UV energy cutoff $E_{\Lambda}$ as follows, 
\begin{align}
    k_{\Lambda}^{(\pm)} =  \frac{1}{2}\sqrt{\frac{(\omega\pm E_{\Lambda})^2 -4 m^2}{v^2 - 2m\kappa}}.
\end{align}
Furthermore, we note that we have used this relation in the main text to define the form factors as a function of the frequency $f^{(l)}(\omega)$. In particular, to determine the frequency where a transition from the $s$-wave SC to $p$-wave SC occurs, we should satisfy, 
\begin{align}
    f^{(0)}(k_{r}(\omega)) = f^{(1)}(k_{r}(\omega)), 
\end{align}
where $f^{(0)}_{\bk} = \left(1+ d_{z,\bk}/d_{\bk} \right)/2$,  $f^{(1)}_{\bk} = vk/d_{\bk}$. From here, we can observe that in order to satisfy this condition, it is desirable to have a positive band curvature $\kappa$ so that $d_{z,\bk}$ and correspondingly $f^{(0)}_{\bk}$ would decrease with the momentum. Therefore, since $f^{(1)}_{\bk}$ increases monotonically with the momentum, this condition can be satisfied with smaller values of the momentum deviation from the valley center. 

Regarding the numerical parameters chosen in the main text we should mention that in Fig.2 the magnitude of $v$ is chosen such that it results in a large value for $k_{r}\simeq 2/a$ which may not be accessible in lattice models. A more realistic parameter regime is obtained by increasing $v$ which reduces the required values of $k_r$ and still captures the same competition between the p-wave and s-wave superconducting states but now with a larger value for the critical coupling $\mathrm{log}(g_{\mathrm{crit}}/m)\simeq 1$ which is has been numerically verified in our simulations. This issue can be also resolved and it does not affect the feasibility results of our paper since the magnitude the critical coupling strength can be decreased by choosing a smaller value for the chemical potential closer to the optimum value of the chemical potential i.e. $\mu=\Gamma_t/2$ as mentioned in the main text. 

Finally, having a finite deviation from the Dirac points, has an additional dynamical effect which helps our proposal indirectly. We note that in our proposal, to develop a superconducting state, it is essential to have a large Coulomb interaction which can occur as a result of screening the Coulomb interaction. But this screening can also lead to large exciton binding energy in TMDs. Therefore, it might appear that exciton formation could compete with the formation of SC. This issue can be avoided by a strong optical pumping of the system well-above the band gap. More precisely, for the formation of excitons, the excited electrons should relax to the bottom of the conduction band. The effective energy associated with inverse relaxation rate through optical phonons is 2\text{ meV}\cite{Brem2018} which is two orders of magnitude smaller than the Rabi frequency being more than 100\text{ meV}. Correspondingly, in our derivation, we ignore the exciton formation processes.

\subsection{Deriving the Equations of Motion}
\setcounter{equation}{0}
\renewcommand{\theequation}{B.\arabic{equation}}

\subsubsection*{Bosonic Bath}
In this section of the appendix, we obtain the equations of motion (EOM) in the presence of a bosonic bath. In general, the bath used in our formalism and its coupling to our system can be described by the following Hamiltonian,
\begin{align}
    & H_{b} = \sum_{\bk} \nu_{\bk} b^{ \dagger}_{\bk} b_{\bk} \\ & 
    H_{s-b}=\sum_{\bk} \lambda_{\bk} \Big(   b_{\bk}  c^{ \dagger}_{c,\bk}  c^{\eta}_{v,\bk} + h.c. \Big).
\end{align}
Using such a bath, leads to a master equation as, 
\begin{align}
    &\partial_t\rho_{s}(t) = -i [H_{s}, \rho_{s} ] + \sum_{\alpha={v,c}} \Gamma_{\alpha}\mathcal{L}[L_{\alpha, \bk}]\rho_s \label{rhoEOMApp}.
\end{align}
where the action of the Lindbladian superoperator $\mathcal{L}$ with a quantum jump operator $L$ on the density matrix $\rho$, is defined as $\mathcal{L}[L]\rho = L \rho L^{\dagger} - \frac{1}{2}\{L^{\dagger}L,\rho\}$. While in general the decay rates depend on the coupling constant $\lambda_{\bk}$ and the density of states of the bath, to simplify our formalism we consider constant decay rates, $\Gamma_v =\Gamma n_B$, and $\Gamma_c = \Gamma (1+n_B)$, where $n_B$ denotes the effective Bose-Einstein population of the bath which we assume is momentum-independent. 

Here, we are interested in the EOM for the occupation probability, $n^{\eta}_{\alpha, \bk}$, the polarization, $\sigma^{\eta}_{\bk}$, and the anomalous pairing, $s^{\eta}_{\bk}= \tr(\rho_s c^{-\eta}_{c,-\textbf{k}}c^{\eta}_{v,\textbf{k}})$. For convenience of labeling we define the notation $n^{\eta}_{\alpha \alpha, \bk}\equiv n^{\eta}_{\alpha, \bk}$, and $n^{\eta}_{cv, \bk}\equiv \sigma^{\eta}_{\bk}$. To derive the EOM for an arbitrary operator $\mathcal{O} = \tr({\rho_s\hatmathO})$ in the Schrodinger equation we use $\partial_t \mathcal{O} = \tr(\hatmathO \partial_t\rho )$.
To write down the EOMs for these quantities we also need the following identities
\begin{align}
    &\tr(\hatmathO[H,\rho]) = \tr([\hatmathO,H]\rho) \\
    &\tr(\hatmathO\mathcal{L}[L]\rho) = \frac{1}{2}\tr([L^{\dagger}, \hatmathO]L\rho) + \frac{1}{2}\tr(L^{\dagger}[\hatmathO,L]\rho)\label{LindbladIdentity}.
\end{align}
We can study the contributions of the Hamiltonian and Lindbladian in the time evolution separately. For the kinetic Hamiltonian part with $H_K = c^{\dagger}_{\alpha,\bk} h_{\alpha \beta, \bk} c^{\eta}_{\beta, \bk}$ we can use the following identities, 
\begin{align}
    \partial_t n^{\eta}_{\mu \nu, \bk}\Big|_{H_K} &= \tr(c^{\eta \dagger}_{\mu, \bk} c^{\eta}_{\nu, \bk}\partial_t\rho)
    = -i \sum_{\alpha=\{v,c\}}(h_{\nu \alpha, \bk} n^{\eta}_{\mu \alpha, \bk} - h_{\alpha \mu, \bk}n^{\eta}_{\alpha \nu, \bk}).
\end{align}
Similarly, for the interband pairing we get, 
\begin{align}
    \partial_t s^{\eta}_{\bk}\Big|_{H_K} &=
    \tr\left(c^{-\eta}_{c,-\textbf{k}}c^{\eta}_{v,\textbf{k}} \partial_t \rho \right)= -i(\epsilon_{c,\bk} + \epsilon_{v,\bk})s^{\eta}_{\bk}. 
\end{align}
We can also compute the commutators of the anomalous pairing and the electron-electron interaction which is
\begin{align}
H_{\mathrm{e-e}}=\sum_{\eta=\pm1;\bk} \left( \Delta_{\bk}^{\eta *} c^{-\eta}_{c,-\bk} c^{\eta}_{v,\bk} + h.c. \right)+\mathrm{const}. 
\end{align}
The corresponding commutator becomes,
\begin{align}
\partial_t s^{\eta}_{\bk}\Big|_{H_{\mathrm{e-e}}} &= - i \Delta_{\bk}(1- n_{v,\bk}^{\eta} - n_{c,-\bk}^{-\eta}).
\end{align}
For the Lindbladian contributions we employ Eq.\eqref{LindbladIdentity} to obtain, 
\begin{align}
    & \partial_t n^{\eta}_{v,\bk}|_{\mathcal{L}} = -\Gamma n_{B} n^{\eta}_{v,\bk} + \Gamma (1 + n_{B}) n^{\eta}_{c,\bk}, \\
    & \partial_t n^{\eta}_{c,\bk}|_{\mathcal{L}} = \Gamma n_{B} n^{\eta}_{v,\bk} - \Gamma(1 + n_{B}) n^{\eta}_{c,\bk}, \\
    & \partial_t \sigma^{\eta}_{\bk}|_{\mathcal{L}} = - \Gamma \left(\frac{1}{2} + n_{B} \right) \sigma^{\eta}_{\bk}, \\
    & \partial_t s^{\eta}_{\bk}|_{\mathcal{L}} = - \Gamma \left( \frac{1}{2} + n_{B} \right)s^{\eta}_{\bk}. 
\end{align}
After combining the Hamiltonian and Lindbladian contributions, we get, 
\begin{align}
& \partial_t n^{\eta}_{v,\bk} = -i(\Omega^{\eta}_{\bk, x} -i \Omega^{\eta}_{\bk, y} )\sigma^{\eta *}_{\bk} + i(\Omega^{\eta}_{\bk, x} +i\Omega^{\eta}_{\bk, y} )\sigma^{\eta}_{ \bk} - i \Delta^{\eta}_{\bk} s^{\eta *}_{\bk} + i \Delta^{\eta*}_{\bk} s^{\eta}_{\bk} - \Gamma n_{B} n^{\eta}_{v,\bk} + \Gamma (1 + n_{B}) n^{\eta}_{c,\bk},  \\
& \partial_t n^{\eta}_{c,\bk} = i (\Omega^{\eta}_{\bk, x} -i\Omega^{\eta}_{\bk, y} )\sigma^{\eta *}_{\bk} - i (\Omega^{\eta}_{\bk, x} +i\Omega^{\eta}_{\bk, y} )\sigma^{\eta}_{\bk} - i \Delta^{-\eta}_{-\bk} s^{-\eta*}_{-\bk} + i \Delta^{-\eta*}_{-\bk} s^{-\eta}_{-\bk} +  \Gamma n_{B} n^{\eta}_{v,\bk} - \Gamma(1 + n_{B}) n^{\eta}_{c,\bk},  \\
& \partial_t \sigma^{\eta}_{\bk} = i \left(\epsilon_{c,\bk} - \epsilon_{v,\bk} - \omega  \right) \sigma^{\eta}_{\bk} -i(\Omega^{\eta}_{\bk, x} -i\Omega^{\eta}_{\bk, y} )  ( n^{\eta}_{c,\bk} - n^{\eta}_{v,\bk}) - \Gamma \left( \frac{1}{2} + n_{B} \right) \sigma^{\eta}_{\bk},  \\
& \partial_t s^{\eta}_{\bk} = -i( \epsilon_{c,\bk} + \epsilon_{v,\bk}) s^{\eta}_{\bk} - i \Delta^{\eta}_{\bk}( 1 -n^{\eta}_{v,\bk} - n^{-\eta}_{c,-\bk} )- \Gamma \left( \frac{1}{2} + n_{B} \right) s^{\eta}_{\bk}. \label{rotatedEOMsBoson} 
\end{align}
Notice that should we want to take exciton formation into account, in the third equation above which is the EOM for the polarization, we need to consider the Hartree-Fock contribution of the electron-electron interaction in the particle-hole channel. This adds a term as $-i \sum^{\eta}_{\bk'} \bar{U}_{\bk \bk'} \sigma^{\eta}_{\bk'}$ on the right-hand side of the third equation above. We can show that in our system exciton formation and the Cooper instability do not compete with each other. Therefore, even in the presence of a finite density of excitons, we can still have a phase transition into a superconducting state. Thus, in our derivation we drop such terms to simplify our analysis. From the third and fourth equations above, we can obtain the anomalous pairing, 
\begin{align}
  s^{\eta}_{\bk}= - \frac{ \Delta^{\eta}_{\bk}(1- n^{\eta}_{v,\bk} - n^{-\eta}_{c,-\bk})}{ \epsilon_{t,\bk} -i \Gamma (\frac{1}{2} + n_B)}.
\end{align}
and the polarization,
\begin{align}
\sigma_{\bk}^{\eta}= \frac{(\Omega^{\eta}_{\bk, x} -i\Omega^{\eta}_{\bk, y} )  ( n^{\eta}_{c,\bk} - n^{\eta}_{v,\bk})}{ \epsilon_{d,\bk}+ i \Gamma (\frac{1}{2} + n_B)}, 
\end{align}
where $\epsilon_{d,\bk}\equiv \epsilon_{c,\bk} - \epsilon_{v,\bk} - \omega $. These relations can be inserted in the first two EOMs in the steady state, 
\begin{align}
& 0 =  \zeta_{\bk}^{\eta} \left( n^{\eta}_{c,\bk} - n^{\eta}_{v,\bk}\right) + \delta^{\eta}_{\bk} \left(1 - n^{\eta}_{v,\bk}  - n^{-\eta}_{c,-\bk} \right) -\gamma_{v} n^{\eta}_{v,\bk} + \gamma_{c} n^{\eta}_{c,\bk}, \\
& 0 = -\zeta^{\eta}_{\bk} \left( n^{\eta}_{c,\bk} - n^{\eta}_{v,\bk}\right) + \delta^{-\eta}_{\bk} \left(1 - n^{-\eta}_{v,-\bk}  - n^{\eta}_{c,\bk} \right) + \gamma_{v} n^{\eta}_{v,\bk} - \gamma_{c} n^{\eta}_{c,\bk}, 
\end{align}
where we have defined 
\begin{align}
    \gamma_{v} \equiv \frac{n_B}{1 + 2n_B},\quad \gamma_{c} \equiv \frac{1 + n_B}{1 + 2n_B}. 
\end{align}
The equations at the two valleys should be solved together. This gives, 
\begin{align}
&  \left(\zeta^{(+)}_{\bk}  + \delta^{(+)}_{\bk} + \gamma_{v} \right) \left(n^{(+)}_{v,\bk} - \frac{1}{2}\right) - \left(\zeta^{(+)}_{\bk}  + \gamma_{c} \right) \left(n^{(+)}_{c,\bk} - \frac{1}{2}\right) + \delta^{(+)}_{\bk} \left( n^{(-)}_{c,-\bk} - \frac{1}{2}\right) =  \frac{\gamma_{c} - \gamma_{v}}{2}, \\
&  \left(\zeta^{(+)}_{\bk}  + \delta^{(-)}_{\bk} + \gamma_{c} \right) \left(n^{(+)}_{c,\bk} - \frac{1}{2} \right) - \left( \zeta^{(+)}_{\bk} + \gamma_{v} \right) \left(n^{(+)}_{v,\bk} - \frac{1}{2} \right) + \delta^{(-)}_{\bk} \left(n^{(-)}_{v,-\bk} - \frac{1}{2}\right) =  \frac{ \gamma_{v} - \gamma_{c} }{2} , \\
& \left(\zeta^{(-)}_{-\bk}  + \delta^{(-)}_{\bk} + \gamma_{v} \right) \left(n^{(-)}_{v,-\bk} - \frac{1}{2} \right) - \left(\zeta^{(-)}_{-\bk}  + \gamma_{c} \right) (n^{(-)}_{c,-\bk} - \frac{1}{2}) + \delta^{(-)}_{\bk} \left(n^{(+)}_{c,\bk} - \frac{1}{2} \right) =  \frac{ \gamma_{c} - \gamma_{v} }{2}, \\
&  \left(\zeta^{(-)}_{-\bk}  + \delta^{(+)}_{\bk} + \gamma_{c} \right) \left(n^{(-)}_{c,-\bk} - \frac{1}{2} \right) - \left( \zeta^{(-)}_{-\bk} + \gamma_{v} \right) \left(n^{(-)}_{v,-\bk} - \frac{1}{2}\right) + \delta^{(+)}_{\bk} \left(n^{(+)}_{v,\bk} - \frac{1}{2} \right) =  \frac{ \gamma_{v} - \gamma_{c} }{2}  .
\end{align}
where the effective Rabi frequency and pairing amplitude are respectively given by, 
\begin{align}
& \zeta^{\eta}_{\bk} = \frac{(\bar{\Omega}_{\bk, x}^{\eta 2} + \bar{\Omega}_{\bk, y}^{\eta 2})}{\epsilon^{2}_{d,\bk}+   (\frac{1}{2} + n_B)^2 \Gamma^{2}}, \\
& \delta^{\eta}_{\bk} =  \frac{|\Delta^{\eta}_{\bk}|^2}{\epsilon^{\eta 2}_{t,\bk}+  (\frac{1}{2} + n_B)^2 \Gamma^{2}}.
\end{align}
The resulting equations can be rewritten in a matrix form,
\begin{align}
    \begin{pmatrix}
    & \zeta^{(+)}_{\bk}  + \delta^{(+)}_{\bk} + \gamma_{v} & - \zeta^{(+)}_{\bk}- \gamma_{c} & 0 & \delta^{(+)}_{\bk} \\
    & -\zeta^{(+)}_{\bk} - \gamma_{v} & \zeta^{(+)}_{\bk}  + \delta^{(-)}_{\bk} + \gamma_{c} & \delta^{(-)}_{\bk} &0 \\
    & 0 & \delta^{(-)}_{\bk} & \zeta^{(-)}_{-\bk}  + \delta^{(-)}_{\bk} + \gamma_{v} & -\zeta^{(-)}_{-\bk}- \gamma_{c} \\
    & \delta^{(+)}_{\bk}  & 0 & - \zeta^{(-)}_{-\bk}-\gamma_{v} & \zeta^{(-)}_{-\bk}  + \delta^{(+)}_{\bk} + \gamma_{c} 
    \end{pmatrix}
    \begin{pmatrix}
    & n^{(+)}_{v,\bk}- \frac{1}{2}\\
    & n^{(+)}_{c,\bk}- \frac{1}{2}\\
    & n^{(-)}_{v,-\bk}- \frac{1}{2}\\
    & n^{(-)}_{c,-\bk}- \frac{1}{2}
    \end{pmatrix}  
    = \frac{1}{2}(\gamma_{c} - \gamma_{v})  \begin{pmatrix}
    & 1 \\ & -1 \\ & 1 \\ &-1
    \end{pmatrix}.
\end{align}
Here, we are mainly interested in studying the onset of the SC phase transition which implies that we can ignore the pairing amplitude in the above equations so that the matrix on the left becomes block diagonal.
In the limit where the effective Rabi frequency around the $-\bK$ valley i.e. $\bar{\Omega}^{(-)}_{\bk}$, is negligible, after using the conservation of the particle densities in the valence and conduction bands of the two valleys separately, ($n^{(\eta)}_{v,\bk} + n^{(\eta)}_{c,\bk}=1$), these probability populations become,
\begin{align}
& n^{(+)}_{v,\bk} = \frac{1}{2} + \frac{\gamma_{c} - \gamma_{v}}{2 (2\zeta^{(+)}_{\bk} + \gamma_{v} + \gamma_{c})}, \\
& n^{(+)}_{c,\bk} = \frac{1}{2} - \frac{\gamma_{c} - \gamma_{v}}{2 (2\zeta^{(+)}_{\bk} + \gamma_{v} + \gamma_{c})}, \\
& n^{(-)}_{v,-\bk} = 1, \\
& n^{(-)}_{c,-\bk} = 0.
\end{align}
Let us further assume that $n_B=0$ which results in $\gamma_{v} =0$, and $\gamma_{c}=1$. In this limit, it is evident that we can have an effective SC population inversion around one of the valleys because, 
\begin{align}
   & 1 - n^{(+)}_{v,\bk} - n^{(-)}_{c,-\bk} \simeq \frac{1}{2} - \frac{1}{2 (2\zeta^{(+)}_{\bk} + 1)} \\
   & 1 - n^{(-)}_{v,-\bk} - n^{(+)}_{c,\bk} \simeq -\frac{1}{2} + \frac{1}{2 (2\zeta^{(+)}_{\bk} + 1)}.
\end{align}
We should hint that in the weak-drive limit, the right-hand side reduces to $+\zeta^{(+)}_{\bk}$ and $-\zeta^{(+)}_{\bk}$. More generally, after defining the interband $\textit{pairing}$ population, $ n^{(\eta)}_{\mathrm{sc},\bk}\equiv 1 - n^{(\eta)}_{v,\bk} - n^{(-\eta)}_{c,-\bk}$ we have, 
\begin{align}
    & n^{(+)}_{\mathrm{sc},\bk}\equiv 1 - n^{(+)}_{v,\bk} - n^{(-)}_{c,-\bk} = \frac{-1}{2(1 + 2n_B)}\left(\frac{1}{2\zeta^{(+)}_{\bk}+1} - \frac{1}{2\zeta^{(-)}_{-\bk}+1} \right),\\
    & n^{(-)}_{\mathrm{sc},\bk}\equiv 1 - n^{(-)}_{v,-\bk} - n^{(+)}_{c,\bk} = \frac{1}{2(1 + 2n_B)}\left(\frac{1}{2\zeta^{(+)}_{\bk}+1} - \frac{1}{2\zeta^{(-)}_{-\bk}+1} \right).
\end{align}
where we have used the fact that at every momentum we have $n^{(+)}_{v,\bk}+n^{(+)}_{c,\bk}=1$. These equations lead to the linearized gap equation in the main text.

Finally, in deriving the final form of the gap equation from the mean-field solution, we note that in dissipative superfluid or superconducting systems, in general it is possible that the condensate attains a time-dependence \cite{Szymanska2006Nonequilibrium}. Let us decompose the total Hamiltonian into its system and system-bath components $H=H_s + H_{s-b}$, where the former has a kinetic and interaction contributions as $H_s=H_k+H_{\mathrm{int}}$. After integrating out the reservoir degrees of freedom the effective Hamiltonian that is obtained for the system’s degrees of freedom is quadratic and non-Hermitian. Therefore, the total quadratic Hamiltonian that is obtained from summing this contribution and $H_k$ would be non-Hermitian, too. Consequently, if we try to use the corresponding non-Hermitian free energy to find saddle point solutions, it would lead to inconsistency as the contribution of the quadratic terms are non-Hermitian while those of the interaction terms are Hermitian. This problem is solved by using the Keldysh action which has a forward and backward temporal contour such that by differentiating between the retarded, advanced and Keldysh Green’s functions, the Hermiticity of the action is always built into it. The detailed Keldysh calculation is explained in Ref \cite{goldstein2015photoinduced}. In the limit where the dissipation rate is small, this results agrees with the modified mean-field approximation. 

\subsubsection*{Fermionic Bath}

Here, we show that we can obtain similar results with a fermionic bath at a fixed temperature $T$, 
\begin{align}
H_{b} = \sum_{\bk, \alpha} \omega_{\alpha}(\bk) b^{\dagger}_{\alpha,\bk} b_{\alpha,\bk}. 
\end{align}
where $\alpha = \{v,c\}$. 
We consider a system-bath coupling which allows exchanging particles between the system and the reservoir, 
\begin{align}
H_{s-b} = \sum_{\bk, \alpha} t_{\alpha}(\bk) \Big[  c^{\eta\dagger}_{\alpha ,\bk}  b_{\alpha,\bk} +  b^{\dagger}_{\alpha,\bk}  c^{\eta}_{\alpha, \bk}  \Big],  \label{sysBathHamil}
\end{align}
Starting with the system-bath coupling term, we assume a thermal Fermi-Dirac distribution for the bath DOF at temperature $T$, so that these DOF can be traced out. After applying the RWA and eliminating the oscillating terms, we arrive at the following master equation for the density matrix of the driven semiconductor:
\begin{equation}\begin{split} 
\partial_t\rho_{s}(t) = -i [H_{s}(t), \rho_{s} ] + \sum_{\bk,  \alpha={v,c}} \Gamma_{\alpha}(\bk)\Big(n^{F}_{\alpha,\bk} \mathcal{L}[c^{\dagger}_{\alpha , \bk}]\rho_{s} + \big(1 -  n^{F}_{\alpha, \bk}\big) \mathcal{L}[c^{\eta}_{\alpha, \bk}]\rho_{s} \Big)\label{AppRhoEOM}, 
\end{split}\end{equation}
where $n^{F}_{\alpha,\bk}=n^F(\epsilon_{\alpha,\bk})$ is the Fermi-Dirac distribution. The decay rates $\Gamma_{\alpha}(\bk)=2 \pi \sum_{a}|t_a|^2 \nu(\epsilon_{\alpha,\textbf{k}}) u^{\alpha*}_{a,\bk} u^{\alpha}_{a,\bk}$ where $\nu(\epsilon)$ represents the density of states of the bath's electrons at energy $\epsilon$. 

For the pairing amplitude between the valence $\alpha =v$ and conduction band $\alpha = c$, this yields, 
\begin{align}
    \partial_t s_{\bk}\Big|_{H} &=
    \tr(c_{c, \bk}^{\eta\dagger}c_{v, \bk}^{\eta\dagger}\partial_t \rho)=i (\epsilon_{c,\bk} + \epsilon_{v,\bk})s_{\bk}. 
\end{align}
For the Lindbladian part we get,
\begin{align}
    \partial_t \hat{\mathcal{O}} 
    = \frac{1}{2}\Gamma_{\alpha \beta}n_a^F\Big( \tr([c^{\eta}_{\alpha, \bk}, \hatmathO]c^{\eta \dagger}_{\beta, \bk}\rho) + \tr(c^{\eta}_{\alpha, \bk}[\hatmathO, c^{\eta \dagger}_{\beta, \bk}]\rho) \Big) + \frac{1}{2}\Gamma_{\alpha \beta}(1 - n^F_a)\Big(\tr([c^{\eta \dagger}_{\alpha, \bk},\hatmathO]c^{\eta}_{\beta, \bk}\rho) + \tr(c^{\eta \dagger}_{\alpha, \bk}[\hatmathO,c^{\eta}_{\beta, \bk}]\rho) \Big), 
\end{align}
where we have used the creation and annihilation operators in the rotating frame. Consequently, we can assume that oscillating terms in the rotating frame can be ignored. This way, we can time average over the Lindbladian which results in considering only the diagonal terms in the above with $\alpha=\beta$. 
\begin{align}
    \partial_t \hat{\mathcal{O}}\Big|_{\mathrm{RW}}= \frac{1}{2}\Gamma_{\alpha \alpha}n^F_{\alpha}\Big( \tr([c^{\eta}_{\alpha, \bk}, \hatmathO]c^{\eta \dagger}_{\alpha, \bk}\rho) + \tr(c^{\eta}_{\alpha, \bk}[\hatmathO, c^{\eta \dagger}_{\alpha, \bk}]\rho) \Big) + \frac{1}{2}\Gamma_{\alpha \alpha}(1 - n^F_{\alpha})\Big(\tr([c^{\eta \dagger}_{\alpha, \bk},\hatmathO]c^{\eta}_{\alpha, \bk}\rho) + \tr(c^{\eta \dagger}_{\alpha, \bk}[\hatmathO,c^{\eta}_{\alpha, \bk}]\rho) \Big). 
\end{align}
Without loss of generality, in the rest of this section, we assume momentum independent dissipation rates and we label its diagonal components by $\Gamma_{\alpha}$. The terms obtained from expanding the right-hand side are similar to the terms obtained in the Bosonic case. The final result of this expansion reads, 
\begin{subequations}
\begin{align}
& \partial_t n^{\eta}_{v,\bk} = -i\left(\Omega^{\eta}_{\bk, x} -i\Omega^{\eta}_{\bk, y} \right)\sigma^{\eta *}_{\bk} + i\left(\Omega^{\eta}_{\bk, x} +i\Omega^{\eta}_{\bk, y} \right)\sigma^{\eta}_{\bk} + i \Delta^{\eta}_{\bk} s^{*}_{\bk} - i \Delta^{\eta*}_{\bk} s_{\bk} -  \Gamma_{v} ( n^{\eta}_{v,\bk} - n^{F}_{v,\bk}),  \\
& \partial_t n^{\eta}_{c,\bk} = i \left(\Omega^{\eta}_{\bk, x} -i\Omega^{\eta}_{\bk, y} \right) \sigma^{\eta *}_{\bk} - i 
\left(\Omega^{\eta}_{\bk, x} +i\Omega^{\eta}_{\bk, y} \right) \sigma^{\eta}_{\bk} + i \Delta^{-\eta}_{-\bk} s^{-\eta *}_{-\bk} - i \Delta_{-\bk}^{-\eta *} s^{-\eta}_{-\bk} -  \Gamma_{c} (n^{\eta}_{c,\bk} - n^{F}_{c,\bk} ),  \\
& \partial_t \sigma^{\eta}_{\bk} = i( \epsilon_{c,\bk} - \epsilon_{v,\bk} - \omega ) \sigma^{\eta}_{\bk} -i\left(\Omega^{\eta}_{\bk, x} -i\Omega^{\eta}_{\bk, y} \right)  \left( n^{\eta}_{c,\bk} - n^{\eta}_{v,\bk} \right) - \frac{1}{2} \left(\Gamma_{c} + \Gamma_{v} \right) \sigma^{\eta}_{\bk} ,  \\
& \partial_t s^{\eta}_{\bk} = -i \left( \epsilon_{c,\bk} + \epsilon_{v,\bk} \right) s^{\eta}_{\bk} - i \Delta^{\eta}_{\bk}\left( n^{\eta}_{v,\bk} + n^{-\eta}_{c,-\bk} - 1 \right) - \frac{1}{2} \left(\Gamma_{c} + \Gamma_{v} \right) s^{\eta}_{\bk}. \label{rotatedEOMs} 
\end{align}
\end{subequations}
where as before we have ignored the terms which are relevant in exciton formation. 
Next, we derive the steady state solution by assuming constant densities and pairing amplitudes in the rotating frame. We start by obtaining the equations for the anomalous pairing, 
\begin{align}
  s^{\eta}_{\bk}= -\frac{ \Delta^{\eta}_{\bk}\left(1- n^{\eta}_{v,\bk} - n^{-\eta}_{c,-\bk} \right)}{ \epsilon_{t,\bk} -\frac{i}{2} \Gamma_{t}}
\end{align}
where we have defined $\epsilon_{t,\bk} =\epsilon_{c,\bk} + \epsilon_{v,\bk}$, and $\Gamma_{t} = \Gamma_{c} + \Gamma_{v}$. In the next step, we consider the EOM for the polarization $\sigma_k$, 
\begin{align}
\sigma^{\eta}_{\bk}= \frac{\left(\Omega^{\eta}_{\bk, x} -i\Omega^{\eta}_{\bk, y} \right)  
\left( n^{\eta}_{c,\bk} - n^{\eta}_{v,\bk}\right)}{ \epsilon_{d,\bk}+ \frac{i}{2} \Gamma_{t}}.
\end{align}
where we have defined $\epsilon_{d, \bk} = \epsilon_{c,\bk}- \epsilon_{v,\bk} - \omega$. Inserting these two equations in the occupation probabilities we get,
\begin{subequations}
\begin{align}
& 0 =  \zeta^{\eta}_{\bk} \left( n^{\eta}_{c,\bk} - n^{\eta}_{v,\bk} \right)+ \delta^{\eta}_{\bk} \left(1 - n^{\eta}_{v,\bk}  - n^{-\eta}_{c,-\bk} \right) -  \gamma_{v} \left( n^{\eta}_{v,\bk} - n^{F}_{v,\bk} \right), \\
& 0 = -\zeta^{\eta}_{\bk} \left( n^{\eta}_{c,\bk} - n^{\eta}_{v,\bk}\right) + \delta^{-\eta}_{\bk}
\left(1 - n^{-\eta}_{v,-\bk}  - n^{\eta}_{c,\bk} \right) -  \gamma_{c} \left( n^{\eta}_{c,\bk} - n^{F}_{c,\bk}\right).
\end{align}
\end{subequations}
where the effective Rabi frequency and the effective pairing amplitudes are, 
\begin{align}
& \zeta^{\eta}_{\bk} = \frac{\bar{\Omega}_{\bk, x}^{\eta 2} + \bar{\Omega}_{\bk, y}^{\eta 2}}{(\epsilon^{2}_{d,\bk}+ \frac{1}{4}\Gamma_t^{2} )} \\
& \delta^{\eta}_{\bk}=\frac{|\Delta^{\eta}_{\bk}|^2}{(\epsilon^{2}_{t, \bk}+ \frac{1}{4}\Gamma_t^{2} )}.
\end{align}
As in the bosonic case, we need to solve four equations simultaneously, 
\begin{subequations}
\begin{align}
&  \left(\zeta^{(+)}_{\bk}  + \delta^{(+)}_{\bk} + \gamma_{v} \right) \left(n^{(+)}_{v,\bk} - \frac{1}{2}\right) - \zeta^{(+)}_{\bk} \left(n^{(+)}_{c,\bk} - \frac{1}{2}\right) + \delta^{(+)}_{\bk} \left(n^{(-)}_{c,-\bk} - \frac{1}{2}\right) =  \gamma_{v} \left(n^{F}_{v,\bk} -\frac{1}{2}\right), \\
&  \left(\zeta^{(+)}_{\bk}  + \delta^{(-)}_{\bk} + \gamma_{c}\right) \left(n^{(+)}_{c,\bk} - \frac{1}{2}\right) - \zeta^{(+)}_{\bk} \left(n^{(+)}_{v,\bk} - \frac{1}{2}\right) + \delta^{(-)}_{\bk} \left(n^{(-)}_{v,-\bk} - \frac{1}{2}\right) =  \gamma_{c} \left(n^{F}_{c,\bk} -\frac{1}{2}\right), \\
&  \left(\zeta^{(-)}_{-\bk}  + \delta^{(-)}_{\bk} + \gamma_{v}\right) \left(n^{(-)}_{v,-\bk} - \frac{1}{2}\right) - \zeta^{(-)}_{-\bk} \left(n^{(-)}_{c,-\bk} - \frac{1}{2}\right) + \delta^{(-)}_{\bk} \left(n^{(+)}_{c,\bk} - \frac{1}{2}\right) =  \gamma_{v} \left(n^{F}_{v,-\bk} -\frac{1}{2}\right), \\
&  \left(\zeta^{(+)}_{\bk}  + \delta^{(+)}_{\bk} + \gamma_{c}\right) \left(n^{(-)}_{c,-\bk} - \frac{1}{2}\right) - \zeta^{(-)}_{-\bk} \left(n^{(-)}_{v,-\bk} - \frac{1}{2}\right) + \delta^{(+)}_{\bk} \left(n^{(+)}_{v,\bk} - \frac{1}{2}\right) =  \gamma_{c} \left(n^{F}_{c,-\bk} -\frac{1}{2}\right).
\end{align}
\end{subequations}
where $\Gamma_t = \Gamma_v + \Gamma_c$, $\gamma_{\alpha} = \Gamma_{\alpha}/\Gamma$. 
We can rewrite these equations in a matrix form, 
\begin{align}
    \begin{pmatrix}
    & \zeta^{(+)}_{\bk}  + \delta^{(+)}_{\bk} + \gamma_{v} & - \zeta^{(+)}_{\bk} & 0 & \delta^{(+)}_{\bk} \\
    & -\zeta^{(+)}_{\bk} & \zeta^{(+)}_{\bk}  + \delta^{(-)}_{\bk} + \gamma_{c} & \delta^{(-)}_{\bk} &0 \\
    & 0 & \delta^{(-)}_{\bk} & \zeta^{(-)}_{-\bk}  + \delta^{(-)}_{\bk} + \gamma_{v} & -\zeta^{(-)}_{-\bk} \\
    & \delta^{(+)}_{\bk}  & 0 & - \zeta^{(-)}_{-\bk} & \zeta^{(-)}_{-\bk}  + \delta^{(+)}_{\bk} + \gamma_{c} 
    \end{pmatrix}
    \begin{pmatrix}
    & n^{(+)}_{v,\bk}- \frac{1}{2}\\
    & n^{(+)}_{c,\bk}- \frac{1}{2}\\
    & n^{(-)}_{v,-\bk}- \frac{1}{2}\\
    & n^{(-)}_{c,-\bk}- \frac{1}{2}
    \end{pmatrix} 
    = \begin{pmatrix}
    & \gamma_{v} (n^{F}_{v,\bk} -\frac{1}{2}) \\
    & \gamma_{c} (n^{F}_{c,\bk} -\frac{1}{2}) \\
    & \gamma_{v} (n^{F}_{v,-\bk} -\frac{1}{2}) \\
    & \gamma_{c} (n^{F}_{c,-\bk} -\frac{1}{2})
    \end{pmatrix}.
\end{align}
In general one needs to invert the matrix on the left to find the solutions for the occupation probabilities. As the first step, we consider the linearized gap equation where we only consider the solutions of the above equation in the zeroth order of $\Delta_{\bk}$. Furthermore, we consider the zero-temperature limit where $n_{\bk}^{F,v/c} = 0, 1$ where $\Delta^{(-)}_{-\bk}=0$ for $\bk$ around the $\bK^{\prime}$ Dirac cone. This yields, 
\begin{align}
& n^{(+)}_{v,\bk} + n^{(-)}_{c,-\bk} - 1 = \frac{\gamma_{v}\gamma_{c} (\gamma_{c} \zeta^{(+)}_{\bk}-\gamma_{v}\zeta^{(-)}_{-\bk}) + (\gamma_{c}^2-\gamma_{v}^2) \zeta^{(-)}_{-\bk} \zeta^{(+)}_{\bk} }{\left(\gamma_{c} \zeta^{(-)}_{-\bk} + \gamma_{v}\gamma_{c} + \gamma_{v}\zeta^{(+)}_{\bk} \right) \left(\gamma_{c} \zeta^{(+)}_{\bk} + \gamma_{v}\gamma_{c} + \gamma_{v}\zeta^{(-)}_{-\bk} \right)} +\mathcal{O}(\Delta^2). \\
& n^{(-)}_{v,-\bk} + n^{(+)}_{c,\bk} - 1 = \frac{\gamma_{v}\gamma_{c} (\gamma_{c} \zeta^{(-)}_{-\bk} - \gamma_{v}\zeta^{(+)}_{\bk}) + (\gamma_{c}^2-\gamma_{v}^2) \zeta^{(-)}_{-\bk} \zeta^{(+)}_{\bk} }{\left(\gamma_{c} \zeta^{(-)}_{-\bk} + \gamma_{v}\gamma_{c} + \gamma_{v}\zeta^{(+)}_{\bk} \right) \left(\gamma_{c} \zeta^{(+)}_{\bk} + \gamma_{v}\gamma_{c} + \gamma_{v}\zeta^{(-)}_{-\bk} \right)} +\mathcal{O}(\Delta^2). 
\end{align}
We can further simplify these relations in the limit that the Rabi frequency around the $K^{\prime}$ point is negligible, 
\begin{align}
& n^{(+)}_{v,\bk} + n^{(-)}_{c,-\bk} - 1 \simeq \frac{-\gamma_{c} \zeta^{(+)}_{\bk}}{ (\gamma_{v}+\gamma_{c}) \zeta^{(+)}_{\bk} + \gamma_{c} \gamma_{v}} + \frac{\zeta^{(-)}_{-\bk}}{\gamma_{c}}, \\
& n^{(-)}_{v,-\bk} + n^{(+)}_{c,\bk} - 1 \simeq \frac{\gamma_{v}\zeta^{(+)}_{\bk}}{ (\gamma_{v}+\gamma_{c}) \zeta^{(+)}_{\bk} + \gamma_{c} \gamma_{v}} - \frac{\zeta^{(-)}_{-\bk}}{\gamma_{v}}.
\end{align}
The above relations can be employed for the interband pairing which can be used to derive the gap equation. To perform this task we need to write the self-consistency definition of mean field order parameter. The result of this calculation yields, 
\begin{align}
\Delta_{\bk}^{\eta} = -\sum_{\bk'}\ \bar{U}_{\bk \bk'} \frac{\epsilon_{t,\bk'}}{\epsilon_{t,\bk'}^{2} + \Gamma_t^{2}  }n^{\eta}_{\mathrm{sc}, \bk}\Delta_{\bk'}^{\eta}, 
\end{align}
where we have used the definition $n^{\eta}_{\mathrm{sc}, \bk}= 1- n^{\eta}_{v,\bk'} - n^{-\eta}_{c,\bk'}$. As in the bosonic bath case, we can see that this equation can be only satisfied around one of the valleys, which for our choice of the laser's polarization will be the $\bK$ valley. Therefore, we can drop the valley index and rewrite this equation as, 
\begin{align}
    \Delta_{\bk} &= -\sum_{\bk'}\ \bar{U}_{\bk \bk'} \frac{\epsilon_{t,\bk'}}{\epsilon_{t,\bk'}^{2} + \Gamma_t^{2}  } \Big( \frac{\gamma_{c} \zeta^{(+)}_{\bk}}{ (\gamma_{v}+\gamma_{c}) \zeta^{(+)}_{\bk} + \gamma_{c} \gamma_{v}} - \frac{\zeta^{(-)}_{-\bk}}{\gamma_{c}} \Big)\Delta_{\bk'}.
\end{align}
Since the only difference of this gap equation and the gap equation in the bosonic bath case is in the effective value of $n^{\eta}_{\mathrm{sc}, \bk}$, we can use the same ansatz for the pairing amplitude as before,
\begin{align}
\Delta^{(l)*}_{\bk} = e^{-i l \phi_{\bk}} f_{\bk}^{(l)}\Delta^{(l)}.
\end{align}
Using this ansatz we can evaluate the critical value of the coupling constant $g$ numerically. After employing the same integration method, we obtain a similar behavior for $g_{\mathrm{crit}}$ as a function of the frequency of the pump, and we observe that a transition from a $s$-wave SC pairing to a $p$-wave pairing is possible. This shows that the phenomenon we observe is due to the specific form of the electron-electron interaction that we engineer and independent of the type of the bath that we use in our model.

\subsection{Chiral edge states}
\setcounter{equation}{0}
\renewcommand{\theequation}{C.\arabic{equation}}

Here, we investigate the possibility of having protected edge states which can carry supercurrent for our system \cite{Qi2011Topological}. 
In particular, we study the possibility of having localized modes in the presence of a hard-wall boundary which is parallel to $\hat{y}$ which requires the wave function to vanish at at $x=0$. 

We assume that we are in the regime where the SC pairing order parameter has acquired a significant value and is no longer negligible as we previously imagined in the course of obtaining the dominant form of the SC order parameter. Besides, we recall that the main effect of the dissipation in our system is to allow the formation of an out-of-equilibrium steady states where pairing is possible for electrons around the resonant ring corresponding to momentum $\bk \approx \bk_r + \delta \bk$ where $|\delta \bk|<|\bk_r|$. Similar to our calculations in the main text, this goal is achievable by a relatively small value of the system-bath coupling. Therefore, for our purpose which is to study the topological properties of the system after reaching this state, we can ignore the system-bath coupling and only consider the Hermitian terms of the effective Hamiltonian in our system. 

In the rotating frame the kinetic energies of the valence and conduction bands around the two valleys are, 
\begin{align}
\epsilon_{\alpha=\{v,c\}, \bk}^{\eta} = \mu + \alpha (d_{\bk} -\frac{\omega}{2}).    
\end{align}
where on the right we set $\alpha = \pm 1$ which corresponds to the conduction and valence bands energies, respectively. Correspondingly, the second-quantized form of the kinetic term reads, $H_{\mathrm{K}} = \sum_{\alpha, \bk} \epsilon_{\alpha, \bk} c_{\alpha, \bk}^{\eta \dagger}c_{\alpha, \bk}^{\eta}.$ Up to quadratic order $d_{\bk} \approx m + \frac{k^2}{2m}(v^2 - 2m \kappa)$.

For states with momentum close to resonant momentum $k_{r} = \frac{1}{2}\sqrt{\frac{\omega^2 -4 m^2}{v^2 - 2m\kappa}}$, we get $d_{\bk_{r}+\delta\bk} \approx \frac{\omega}{2} + \frac{\bk_{r}.\delta \bk }{m}(v^2 - 2m \kappa)$. Using this approximation the kinetic energies read, 
\begin{align}
\epsilon_{\alpha=\{v,c\}, \bk}^{\eta} = \mu + \alpha\frac{\tilde{v}^2}{m}\bk_{r}.\delta \bk , 
\end{align}
where $\tilde{v}^2 = v^2 - 2m \kappa$. The light induced modification of the band structure is implemented through the Rabi vectors in the two valleys $\bOmega^{\eta}_{\bk}$ as found in Eq.\eqref{rwOmega} and \eqref{rwOmegaTilde}. To simplify our study, we consider $k_y = 0$. The resonance surface is reduced to the resonance points at $\bk_{r} = \pm k_r \hat{x}$. Without loss of generality we also set the band  curvature to zero ($\kappa = 0$). The magnitude of the Rabi vectors then reads as, 
\begin{align}
    \Omega_{k, x}^{(+)} =\left( 1 - \frac{v^2 (k_r^2 + 2 k_{r,x} \delta k_x)}{4m^2} \right)\Omega_0 , \quad     \Omega_{k, x}^{(-)} = \frac{v^2 }{4m^2}  (k_r^2 + 2 k_{r,x} \delta k_x)\Omega_0.\label{OmegaKappa0}
\end{align}
where $\Omega_0 = eA_0 v$.

We note that after thermalization, the pairing obtains a finite value in the basis above where the kinetic energies are diagonalized. Thus as the next step, we add the BdG pairing Hamiltonian of the system, to the Hamiltonian above. 

Next, we transform the pairing Hamiltonian of the system in the momentum space in the mean field limit where $\Delta_{\bk}^{(-)} \equiv \Delta_{\bk}$ is finite and $\Delta_{\bk}^{(+)}$ is vanishing. The pairing Hamiltonian is 
\begin{align}
    H_{\mathrm{SC}} &= \sum_{\bk} \Delta_{\bk} c^{(-)\dagger}_{v,  +\bk}c^{(+)\dagger}_{c,-\bk} + \mathrm{h.c.} 
\end{align}
where superconducting order parameter $\Delta_{\bk}=\Delta_0 (k_x \pm ik_y)$ correspond to a chiral $p$-wave. As outlined above, we initially consider $k_y=0$. 

The sum of the kinetic and pairing Hamiltonian can be combined in a BdG form, $H_{\mathrm{tot}} =\sum_{\bk} \Psi_{\bk}^{\dagger}\mathcal{H}_{\mathrm{tot}} \Psi_{\bk} $, by using the four-component spinor $\Psi_{\bk}^{\dagger}= \begin{pmatrix} c_{v, \bk}^{(-)\dagger} & c_{c, -\bk}^{(+)} & c_{v, -\bk}^{(+)} & c_{c, \bk}^{(-)\dagger}\end{pmatrix}$. The corresponding first-quantized Hamiltonian has the form: 
\begin{align}
\mathcal{H}_{\mathrm{tot}} = \begin{pmatrix} \epsilon_{v,\bk} &\Delta_{\bk} & 0 & \Omega_{\bk,x}^{(-)}-i\Omega_{\bk,y}^{(-)} \\
\Delta_{\bk}^* & -\epsilon_{c,\bk} & -\Omega_{-\bk, x}^{(+)} - i \Omega_{-\bk, y}^{(+)} & 0 \\ 0 & -\Omega_{-\bk, x}^{(+)} + i \Omega_{-\bk, y}^{(+)} & -\epsilon_{v, \bk} & 0 \\
\Omega_{\bk,x}^{(-)}+i\Omega_{\bk,y}^{(-)}& 0 & 0 & \epsilon_{c,\bk}
\end{pmatrix}.
\end{align}
We first consider the limit where the pairing amplitude is vanishing in the Hamiltonian above, $ \mathcal{H}_{\mathrm{kin}}=\mathcal{H}_{\mathrm{SC}}|_{\Delta =0}$, 
\begin{align}
\mathcal{H}_{\mathrm{kin}} = \begin{pmatrix} \epsilon_{v,\bk} & 0 & 0 & \Omega_{\bk,x}^{(-)}-i\Omega_{\bk,y}^{(-)} \\
0 & -\epsilon_{c,\bk} & -\Omega_{-\bk, x}^{(+)} - i \Omega_{-\bk, y}^{(+)} & 0 \\ 0 & -\Omega_{-\bk, x}^{(+)} + i \Omega_{-\bk, y}^{(+)} & -\epsilon_{v, \bk} & 0 \\
\Omega_{\bk,x}^{(-)}+i\Omega_{\bk,y}^{(-)}& 0 & 0 & \epsilon_{c,\bk}
\end{pmatrix}.
\end{align}
Due to the block-diagonal structure of the Hamiltonian, we can diagonalize it by applying a simple rotation. The eigenenergies at the two valleys are, 
\begin{align}
\tilde{\epsilon}^{\eta}_{\alpha, \bk} = \epsilon_{\alpha,\bk} \pm \left(\epsilon_{\alpha,\bk}^2 + \bOmega_{\bk}^{\eta 2} \right)^{1/2}.
\end{align}
which around the resonance surface can be approximated as, $\tilde{\epsilon}^{\eta}_{\alpha, \bk} \simeq \epsilon_{\alpha,\bk} \pm \Omega_{\bk}^{\eta 2}$. Thus we define, 
\begin{align}
    E^{(1,2)} = \mu \pm \Omega^{(+)}_{\bk}, \quad E^{(3, 4)} = \mu \pm \Omega^{(-)}_{\bk}.
\end{align}
The required rotation can be obtained by finding the eigenstates of the above Hamiltonian around the resonant region,
\begin{align}
    |\phi^{(1)}_{\bk}\rangle= \frac{1}{\sqrt{2}} \begin{pmatrix}
    1 \\ 0 \\ 0 \\ e^{i\theta_{\Omega^{+}_{\bk}}}
    \end{pmatrix}, \quad
    |\phi^{(1)}_{\bk}\rangle= \frac{1}{\sqrt{2}} \begin{pmatrix}
    1 \\ 0 \\ 0 \\ -e^{i\theta_{\Omega^{+}_{\bk}}}
    \end{pmatrix},\quad 
    |\phi^{(3)}_{\bk}\rangle= \frac{1}{\sqrt{2}} \begin{pmatrix}
    1 \\ 0 \\ 0 \\ e^{i\theta_{\Omega^{-}_{\bk}}}
    \end{pmatrix},   \quad
|\phi^{(4)}_{\bk}\rangle= \frac{1}{\sqrt{2}} \begin{pmatrix}
    1 \\ 0 \\ 0 \\ -e^{i\theta_{\Omega^{-}_{\bk}}}
    \end{pmatrix}.    
\end{align}
where we have defined $\theta^{\eta}_{\bk} = \tan^{(-1)}(\Omega^{\eta}_{\bk, y}/\Omega^{\eta}_{\bk, x})$ and used $\Omega^{\eta}_{-\bk}=\Omega^{\eta}_{\bk}$. Now, we can rewrite the original Hamiltonian, $\mathcal{H}_{\mathrm{SC}}$ in this rotated basis, 
\begin{align}
    \tilde{\mathcal{H}}_{\mathrm{tot}} = \begin{pmatrix} \mu + \Omega^{(+)}_{\bk} & 0 & \Delta_{\bk}/2 &\Delta_{\bk}/2 \\
    0 & \mu - \Omega^{(+)}_{\bk}  & \Delta_{\bk}/2 & \Delta_{\bk}/2 \\ \Delta_{\bk}^*/2 & \Delta_{\bk}^*/2 & -\mu - \Omega^{(-)}_{\bk} & 0 \\
\Delta_{\bk}^*/2 & \Delta_{\bk}^*/2 & 0 & -\mu + \Omega^{(-)}_{\bk}
\end{pmatrix}.
\end{align}
We can rewrite this Hamiltonian in a more compressed way by introducing the Pauli matrices for the intra and inter valley matrix elements respectively denoted by $\sigma_{i}, \tau_{i}$ where $i = \{0, x, y, z\}$. Using this notation the rotated Hamiltonian becomes, 
\begin{align}
    \tilde{\mathcal{H}}_{\mathrm{tot}} = \frac{\Delta_{\bk}}{2}(\sigma_x+\sigma_0)\tau_x + \mu \sigma_0 \tau_z + \frac{\Omega^{(+)}_{\bk} + \Omega^{(-)}_{\bk}}{2}\sigma_z\tau_z + \frac{\Omega^{(+)}_{\bk} - \Omega^{(-)}_{\bk}}{2}\sigma_z\tau_0.
\end{align}

{Since this Hamiltonian is a $4\times 4$ matrix which is difficult to analytically diagonalize, we start by studying a modified version of this Hamilotnian where the $\frac{\Delta_{\bk}}{2}\sigma_0\tau_x$ term is vanishing. As we will show later this is viable because this term  Hamiltonian hosts protected topological edge states when the pairing has a chiral p-wave structure. In the presence of this term, in general the edge states can hybridize with the bulk states which are away from the resonance points. However, since the occupation of Cooper pairs is negligible away from the resonance points, in our non-equilibrium setting the hybridization of bulk states will be negligible. Hence, in the following we consider the following Hamiltonian, }

\begin{align}
    \mathcal{H}_{\mathrm{eff}} = \frac{\Delta_{\bk}}{2} \sigma_x \tau_x + \mu \sigma_0 \tau_z + \frac{\Omega^{(+)}_{\bk} + \Omega^{(-)}_{\bk}}{2}\sigma_z\tau_z + \frac{\Omega^{(+)}_{\bk} - \Omega^{(-)}_{\bk}}{2}\sigma_z\tau_0.
\end{align}
We can easily verify that since this Hamiltonian commutes with the matrix $\sigma_z \tau_z$, the two matrices can be simultaneously diagonlized. The latter matrix has eigenvalues of $\pm1$, and therefore, we can diagonlize $\mathcal{H}_{\mathrm{eff}}$ in the $\pm 1$ sectors of $\sigma_z \tau_z$. 
\setlist[itemize]{align=parleft,left=0pt..0em}
\begin{itemize}
\item \textbf{\ \ +1 Sector:}

Let us first consider the $+1$ sector by inserting $k_{x} = k_{r, x}+\delta k_x$. To translate the momentum-space Hamiltonian to the real-space we insert $\delta k_x \rightarrow -i \partial_x$. Here, we can introduce the following relevant eigenstates: $|1\rangle \equiv |\sigma_z =1, \tau_z = 1\rangle $, and $|2\rangle \equiv |\sigma_z = -1, \tau_z = -1\rangle$. To simplify our notation, we introduce the Pauli matrices $\xi_i $ in the space of $|1, 2\rangle$ states. After using Eq.\eqref{OmegaKappa0} the resulting Hamiltonian becomes, 
\begin{align}
    \mathcal{H}_{\mathrm{eff}} &= \mu \xi_z  + \frac{\Omega_0}{2}\xi_0 + \left(1 - \frac{v^2 k_r^2}{2m^2}\right) \Omega_0 \xi_z -  \frac{v^2 k_{r, x} \Omega_0 \delta k_x}{m^2}\xi_z +\frac{\Delta_{\bk_r}}{2} \xi_x \notag \\
    &= \frac{\Omega_0}{2}\xi_0 + \left[ \mu  + \left(1 - \frac{v^2 k_r^2}{2m^2}  +  \frac{i v^2 k_{r, x} \partial_x}{m^2} \right)\Omega_0  \right] \xi_z + \frac{\Delta_{\bk_r}}{2} \xi_x.\label{HVolPlusSec}
\end{align}
Let us denote the eigenstates of this Hamiltonian by $|\psi\rangle$. Next, we apply a gauge transformation $|\psi\rangle \rightarrow |\psi'\rangle = e^{-i \gamma_{\bk_r} x} |\psi\rangle$ where we choose $\gamma$ such that the constant terms in the bracket above will cancel each other. This gives,
\begin{align}
    \gamma_{\bk} = -m^2 \frac{\mu  + \left(1 - \frac{v^2 k_r^2}{2m^2}\right)\Omega_0}{v^2 k_{r, x} \Omega_0}.
\end{align}
After this insertion our Hamiltonian becomes, 
\begin{align}
    \mathcal{H'}_{\mathrm{eff}} &= \Omega_0 \xi_z + \frac{i v^2 k_{r, x}\partial_x}{m^2} \Omega_0 \xi_z + \frac{\Delta_{\bk_r}}{2} \xi_x.
\end{align}
This Hamiltonian has a localized eigenstates given by, 
\begin{align}
     |\psi'(x)\rangle= e^{\lambda_{\bk_r} x}|\eta_y = -1 \rangle,\quad \lambda_{\bk_r}  = \frac{\Delta_{\bk_r}}{k_{r, x}}\frac{m^2}{v^2 \Omega_0}. 
\end{align}
Since this eigenstate vanishes at $x \rightarrow \infty$, it satisfies the boundary condition for a semi-infinite strip geometry with a boundary at $x=0$ and extended along $x\rightarrow -\infty$. 

In a geometry where we impose hard-wall boundary conditions, the wave function should vanish at $x=0$. Hence, under such conditions the wave functions with momenta around the two resonance points $k_x = k_r$ and $k_x = -k_r$ superpose. To obtain the localized wave functions with momentum around $\bk = -k_r\hat{x}$, we notice that due to the odd parity of the SC order parameter, $\Delta_{-\bk_r} = -\Delta_{\bk_r}$, we have $\lambda_{-\bk_r}=-\lambda_{\bk_r}$ and $\gamma'_{-\bk} = \gamma'_{\bk}$. More importantly, we should note that the spinor of the localized state at $-\bk_r$ is the same as the spinor of the state at $\bk_r$, namely $|\xi_y = -1\rangle$. Thus, we can build a superposition of the states with opposite momenta $\bk_r$ and $-\bk_r$, to form a state which vanishes at $x=0$, decays exponentially for $x\rightarrow - \infty$ according to,
\begin{align}
    |\psi \rangle &= \sin(\gamma_{\bk_r}x)e^{\lambda_{\bk_r}x}|\xi_y =-1\rangle\notag \\
    &=\sin(\gamma_{\bk_r}x)e^{\lambda_{\bk_r}x}\left(|\sigma_x =1, \tau_x = 1\rangle  - i |\sigma_x =-1, \tau_x = -1\rangle \right), 
\end{align}
and whose energy is, $E_{\psi} = \Omega_0/2$. This state is the only localized state on the $x=0$ boundary, which originates from the $+1$ sector which corresponds to $\tau_x \sigma_x = 1$. 
\item \textbf{\ \ -1 Sector:}

In the other sector we project the Hamiltonian to the subspace where $\tau_x \sigma_x = -1$. This subspace is spanned by the following states, 
$|3\rangle = |\sigma_z=1, \tau_z = -1\rangle,$ and $
|4\rangle = |\sigma_z=-1, \tau_z = 1\rangle$, where as before we use $\xi_i$'s to denote the Pauli matrices in this basis. Consequently, the Hamiltonian required for the Volovik's approach becomes, 
\begin{align}
    \mathcal{H}_{\mathrm{eff}} &= -\mu \xi_z  - \frac{\Omega_0}{2}\xi_0 + \left(1 - \frac{v^2 k_r^2}{2m^2}\right) \Omega_0 \xi_z + \frac{v^2 k_{r, x} \Omega_0 \delta k_x}{m^2}\xi_z +\frac{\Delta_{\bk_r}}{2} \xi_x \notag \\
    &= -\frac{\Omega_0}{2}\xi_0 + \left[ -\mu  + \left(1 - \frac{v^2 k_r^2}{2m^2}  -  \frac{i v^2 k_{r, x} \partial_x}{m^2} \right)\Omega_0  \right] \xi_z + \frac{\Delta_{\bk_r}}{2} \xi_x.    
\end{align}
which should be compared with Eq.\eqref{HVolPlusSec}. As before we apply a phase shift to the wave function as $|\psi\rangle \rightarrow |\psi'\rangle=e^{-i\gamma_{\bk_r}x}|\psi\rangle$. The required phase is given by, 
\begin{align}
    \gamma'_{\bk_r} = m^2 \frac{-\mu  + \left(1 - \frac{v^2 k_r^2}{2m^2}\right)\Omega_0}{v^2 k_{r, x} \Omega_0}
\end{align}
Similar to the $+1$ sector, we can find a localized state around $x=0$, by superposing the localized states in the vicinity of $\bk_r$ and $-\bk_r$. The final resulting state is, 
\begin{align}
    |\psi'\rangle &=\sin(\gamma'_{\bk_r}x)e^{\lambda_{\bk_r} x} |\xi_y =1\rangle \notag \\
    &= \sin(\gamma'_{\bk_r}x)e^{\lambda_{\bk_r} x}\left( |\sigma_z =1, \tau_z =-1\rangle + i |\sigma_z =-1, \tau_z =1\rangle\right). 
\end{align}
whose energy is $-\Omega_0/2$. 
\end{itemize}
The analysis above demonstrates that in the absence of the matrix $\tau_x \sigma_0$ in the Hamiltonian $\mathcal{H}_{\mathrm{eff}}$, we have two chiral localized states on the edge with energies $\pm \Omega_0$. To show that the states are chiral with opposite chirality we need to consider small momentum $k_y$ parallel to the $x=0$ edge. Such small momentum corresponds to the addition of a term of the form $\Delta_0 k_y \tau_x \sigma_y$. On can readily see that the edge states $|\psi\rangle$ and $|\psi'\rangle$ are eigenstates of $k_y \tau_x \sigma_y$ with eigenvalue $\pm 1$ correspondingly. As a result, they disperse linearly with parallel momentum with opposite velocities and form oppositely chiral supercurrents.

Now the presence of pairing elements associated with $\tau_x \sigma_0$, in general can mix the corresponding eigen states in the two sectors. This can be easily shown by evaluating the matrix elements of $\tau_x \sigma_0$ between the localized states and continuum states in the two sectors. For large system sizes due to the exponential decay of the localized states at the boundaries, the hybridization of the localized states from different sectors is negligible. However, we should still consider the possibility of the hybridization of the localized states of one sector with the bulk states of the other sectors. To answer this question, let us first find the functional form of the bulk states. Here, we only consider the $+1$-sector and the bulk states of the other sector can be obtained in a similar manner. The eigenvalues are conveniently obtained by diagonalizing the corresponding Hamiltonian which (for the two sectors) gives, 
\begin{eqnarray}
    E^{(+1)} &=&  \frac{\Omega_0}{2}\pm \sqrt{\Delta_{\bk}^2 + \left(\frac{v^2 k_r \Omega_0}{m^2}\right)^2\delta k_x^2} \\
    E^{(-1)} &=&  -\frac{\Omega_0}{2}\pm \sqrt{\Delta_{\bk}^2 + \left(\frac{v^2 k_r \Omega_0}{m^2}\right)^2\delta k_x^2} 
\end{eqnarray}
The associated eigenstates are: 

\begin{eqnarray}
|\Psi\left(\delta k_x\right)\rangle &=&\sin(\gamma_{\bk_r}x)\left(\left[\epsilon_{\delta k_x}+\sqrt{v^2}{m^2}k_r \Omega_0\right] |\sigma_x =1, \tau_x = 1\rangle \pm \Delta_{\bk} |\sigma_x =-1, \tau_x = -1\rangle \right)/\sqrt{\mathcal{N}} \\ |\Psi'\left(\delta k_x\right)\rangle &=&\sin(\gamma'_{\bk_r}x)\left(\left[\epsilon_{\delta k_x}+\sqrt{v^2}{m^2}k_r \Omega_0\right] |\sigma_x =1, \tau_x = -1\rangle \pm \Delta_{\bk} |\sigma_x =-1, \tau_x = 1\rangle \right)/\sqrt{\mathcal{N}}.
\end{eqnarray}
where $\mathcal{N}=\left[\epsilon_{\delta k_x}+\sqrt{v^2}{m^2}k_r \Omega_0\right]^2 +\Delta_{\bk}^2$ is the normalization factor. As discussed above $\tau_x\sigma_0$ mixes the two sectors (both for localized and continuum states). Since $\gamma$ and $\gamma'$ for non-zero $\mu$ and $k_r$, are independent, the matrix elements between states from the two sector (localized or continuum) vanishes. As a result, the projection of $\tau_x\sigma_0$ term into the states close to the resonance vanishes and these terms do not contribute to the effective Hamiltonian close to resonant momenta.

However, as mentioned previously, these matrix elements are negligible because in order to match the traveling component of the corresponding localized states and bulk states with  different  momenta,  we  need  to  incorporate  states  with momenta  largely  different  from  the  resonant  momentum.Consequently,  since  such  states  do  not  contribute  to  superconductivity, their corresponding matrix elements can be ignored.

\end{widetext}

\end{document}